\documentclass[twocolumn]{aastex62}
\newcommand \redColor{}
\newcommand \bColor{}
\newcommand \oColor{}
\newcommand \oColorS{}
\renewcommand\labelenumi{(\roman{enumi})}
\renewcommand\theenumi\labelenumi
\renewcommand{\vec}[1]{\boldsymbol{#1}}

\usepackage{amsmath,amsthm,amsfonts,amssymb,bm}
\usepackage{listings}
\usepackage{multirow}
\usepackage{color}

\usepackage{textcomp}
\usepackage{epstopdf}
\usepackage{natbib}
\hypersetup{breaklinks}

\begin{document}
\title{FPFS Shear Estimator: Systematic Tests on the Hyper Suprime-Cam Survey First Year Data}
\author{Xiangchong Li}
\affiliation{Department of Physics, University of Tokyo, Tokyo 113-0033, Japan}
\affiliation{Kavli Institute for the Physics and Mathematics of the Universe (WPI),\\
University of Tokyo, Kashiwa 277-8583, Japan}
\author{Masamune Oguri}
\affiliation{Department of Physics, University of Tokyo, Tokyo 113-0033, Japan}
\affiliation{Kavli Institute for the Physics and Mathematics of the Universe (WPI),\\
University of Tokyo, Kashiwa 277-8583, Japan}
\affiliation{Research Center for the Early Universe, University of Tokyo, Tokyo 113-0033, Japan}
\author{Nobuhiko Katayama}
\affiliation{Kavli Institute for the Physics and Mathematics of the Universe (WPI),\\
University of Tokyo, Kashiwa 277-8583, Japan}
\author{Wentao Luo}
\affiliation{Kavli Institute for the Physics and Mathematics of the Universe (WPI),\\
University of Tokyo, Kashiwa 277-8583, Japan}
\author{Wenting Wang}
\affiliation{Kavli Institute for the Physics and Mathematics of the Universe (WPI),\\
University of Tokyo, Kashiwa 277-8583, Japan}
\author{Jiaxin Han}
\affiliation{Department of Astronomy, School of Physics and Astronomy, Shanghai Jiao Tong University, Shanghai, 200240, China}
\affiliation{Kavli Institute for the Physics and Mathematics of the Universe (WPI),\\
University of Tokyo, Kashiwa 277-8583, Japan}
\author{Hironao Miyatake}
\affiliation{Institute for Advanced Research, Nagoya University, Furo-cho, Nagoya 464-8601, Japan}
\affiliation{Division of Particle and Astrophysical Science, Graduate School of Science,\\
Nagoya University, Furo-cho, Nagoya 464-8602, Japan}
\affiliation{Kavli Institute for the Physics and Mathematics of the Universe (WPI),\\
University of Tokyo, Kashiwa 277-8583, Japan}
\author{Keigo Nakamura}
\affiliation{Kavli Institute for the Physics and Mathematics of the Universe (WPI),\\
University of Tokyo, Kashiwa 277-8583, Japan}
\author{Surhud More}
\affiliation{The Inter-University Center for Astronomy and Astrophysics (IUCAA),\\
    Post Bag 4, Ganeshkhind, Pune 411007, India.}
\affiliation{Kavli Institute for the Physics and Mathematics of the Universe (WPI),\\
    University of Tokyo, Kashiwa 277-8583, Japan}

\noaffiliation
\email{xiangchong.li@ipmu.jp}

\begin{abstract}
We apply the Fourier Power Function Shapelets (FPFS) shear estimator to the first year data of the Hyper Suprime-Cam survey to construct a shape catalog. The FPFS shear estimator has been demonstrated to have multiplicative bias less than $1\%$ in the absence of blending, regardless of complexities of galaxy shapes, smears of point spread functions (PSFs) and contamination from noise. The blending bias is calibrated with realistic image simulations, which include the impact of neighboring objects, using the COSMOS Hubble Space Telescope images. Here we carefully test the influence of PSF model residual on the FPFS shear estimation and the uncertainties in the shear calibration.  Internal null tests are conducted to characterize potential systematics in the FPFS shape catalog and the results are compared with those measured using a catalog where the shapes were estimated using the re-Gaussianization algorithms. Furthermore, we compare various weak lensing measurements between the FPFS shape catalog and the re-Gaussianization shape catalog and conclude that the weak lensing measurements between these two shape catalogs are consistent with each other within the statistical uncertainty.
\end{abstract}

\section{Introduction}

Weak lensing provides us with a means of observing the total matter distributions in the universe, including invisible dark matter, by measuring coherent shear distortions on background galaxy images caused by inhomogeneous density distributions along the line-of-sight \citep[e.g.][]{revKilbinger15,revRachel17}.

Weak lensing has wide applications in cosmology. For instance, the cross-correlation between shear measured from background galaxies and positions of foreground lens galaxies, which is commonly known as galaxy-galaxy lensing, probes into the connection between galaxies and underlying matter fluctuations \citep[e.g.][]{gglens-SDSS-Rachel2006,gglens-GAMA-Han2014}. Combined with galaxy clustering, galaxy-galaxy lensing can be used to constrain cosmology \citep[e.g.][]{gglens-BossCFHTMore2015,gglens-DES1}.
The two-point autocorrelation of shear, which is referred to as cosmic shear, directly measures the amplitude and growth of matter fluctuations and hence can be used to constrain cosmology \citep[e.g.][]{cosmicShearRealKids450,cosmicShear-DES1,cosmicShear_HSC1_Fourier2019,cosmicShear_HSC1_configuration2019}.
One can also directly reconstruct projected mass distribution from an observed shear field \citep[e.g.][]{HSC1-massMaps,DES-SV-massMap-sparsity}. Such weak lensing mass maps provide an important means of studying the non-Gaussian features of the matter distributions \citep[e.g.][]{massMap_KIDS450_prediction2018,massMap_KIDS450_inference2018}.

Given its importance, weak lensing is one of the primary science goals of the following three ongoing stage-III surveys, the Kilo-Degree Survey\footnote{\url{http://kids.strw.leidenuniv.nl/index.php}} \citep[KiDS,][]{KIDS13}, the Dark Energy Survey\footnote{\url{http://www.darkenergysurvey.org/}} \citep[DES,][]{DES16} and the Subaru Hyper Suprime-Cam survey\footnote{\url{http://hsc.mtk.nao.ac.jp/ssp/}} \citep[HSC,][]{HSC1-data}.
HSC is a wide-field prime focus camera with a $1.5$ deg diameter field-of-view mounted on the $8.2$-meter Subaru telescope \citep{HSC-HW-Furusawa2018,HSC-HW-Kawanomoto2018,HSC-HW-Komiyama2018,HSC-HW-Miyazaki2018} .
HSC survey has a median $i$-band seeing of $0.6''$ and an $i$-band limiting magnitude of $m_i\sim 26$ for its Wide layer. Such good seeing and deep image enable us to use a large number of galaxies up to $z\sim 2$ for the weak lensing analysis. The Wide layer of the HSC first year data covers $\sim 170 ~\rm{deg}^2$, if we restrict ourselves to the full depth and full color regions without considering the bright star masks \citep[see][]{HSC1-catalog}.

{\oColor{
Accurate shear estimation is a challenging task in weak lesning surveys.
Several traditional methods (eg. \citet[KSB]{KSB95}, \citet[re-Gaussianization]{reGaussianization}) have been proposed to recover shear from large ensemble of galaxies.
However, these traditional methods suffer from noise contamination (noise bias) and irregular galaxy shapes (galaxy model bias). Therefore, realistic image simulations are used to calibrate the noise bias and model bias for these traditional method.
Several newly developed methods (eg. \citet[METACALIBRATION]{metacal2}, \citet[BFD]{BFD16}, \citet[FourierQuard]{Z17}, \citet[FPFS]{Li18FPFS} )can accurately recover shear with good accuracy for isolated galaxies in the absence of blending.
However, for deep ground-based surveys, biases from blending effect should be carefully revised.
METADETECTION\citep{metaDet2019}, which is developed based on METACALIBRATION, offers a solution to biases caused by blending without relying calibration from external simulations.
}}

In this paper, we apply the FPFS shear estimator, which was recently developed in \citet{Li18FPFS}, to the $i$-band images of the  HSC first year data.
\citet{Li18FPFS} proposed the FPFS shear estimator for an accurate shear measurement from large ensembles of galaxy images. The FPFS method conducts the shape measurement on the power function of galaxy image's Fourier transform. It projects the Fourier power function of galaxies on shapelet basis vectors after deconvoling the Point Spread Function (PSF) in Fourier space. Four shapelet modes are used to construct ellipticity and the corresponding shear response.
\begin{figure}
	\centering
	\includegraphics[width=.45\textwidth]{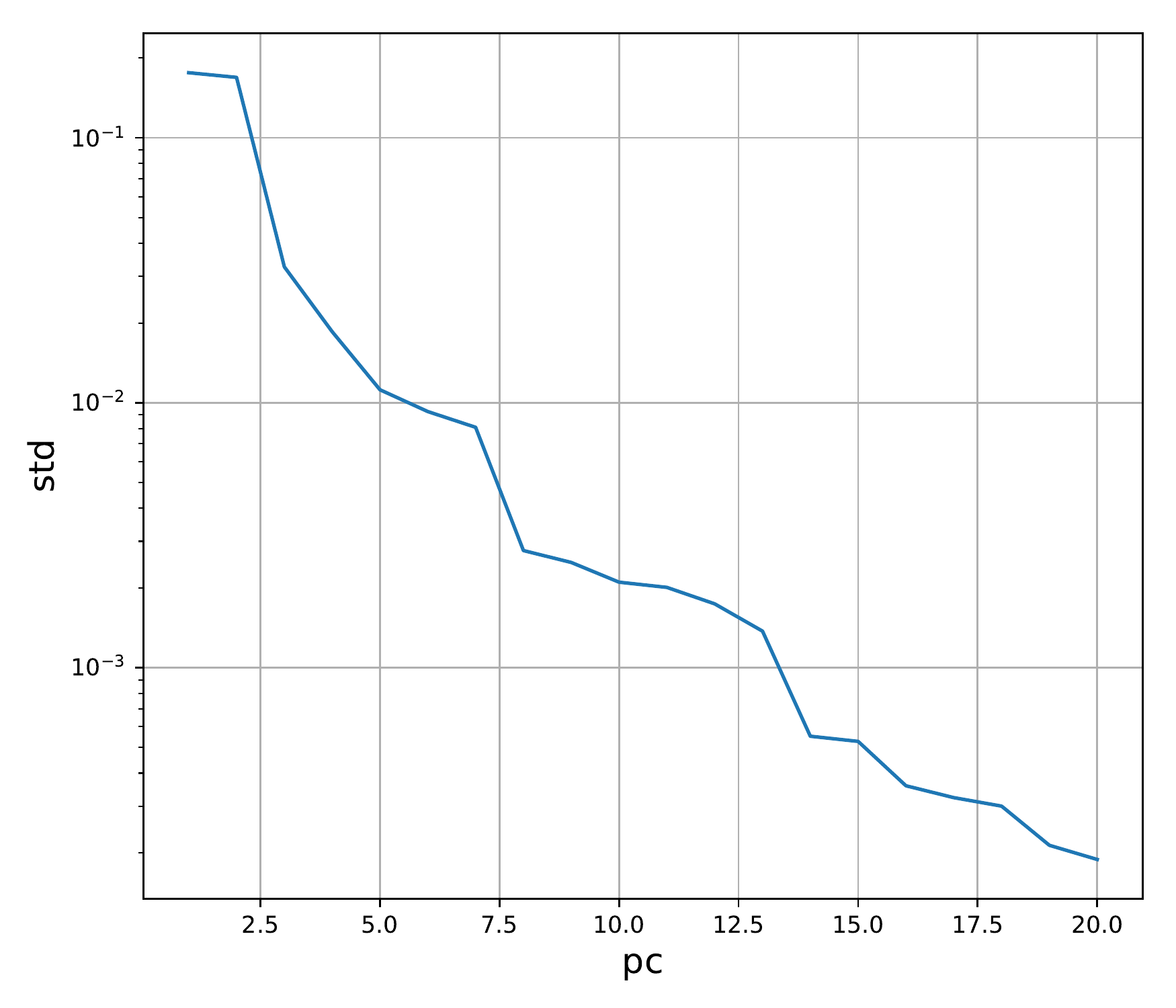}
	\caption{The standard deviation of noise correlation function for different principal components (PCs). The $x$-axis is the rank of PCs and the $y$-axis is the standard deviation of the noise correlation functions for the corresponding PC.} \label{fig_noiCorPCsVar}
\end{figure}
\citet{Li18FPFS} used HSC-like galaxy image simulations from \citet{HSC1-GREAT3Sim} to show that, in the absence of blending, the systematic biases for the FPFS shear estimator are well below $1\%$, regardless of complexities of galaxy morphologies, smears of PSFs and contaminations from noise. However, in the existence of blending, the first generation of HSC deblender \citep{HSC1-pipeline} is used to isolate blended sources before shape measurement and a multiplicative bias of $-5.7\%$ on average is found. Such blending bias is modeled and calibrated with the help of HSC-like image simulations \citep{Li18FPFS}.
\begin{figure}
	\centering
	\includegraphics[width=.45\textwidth]{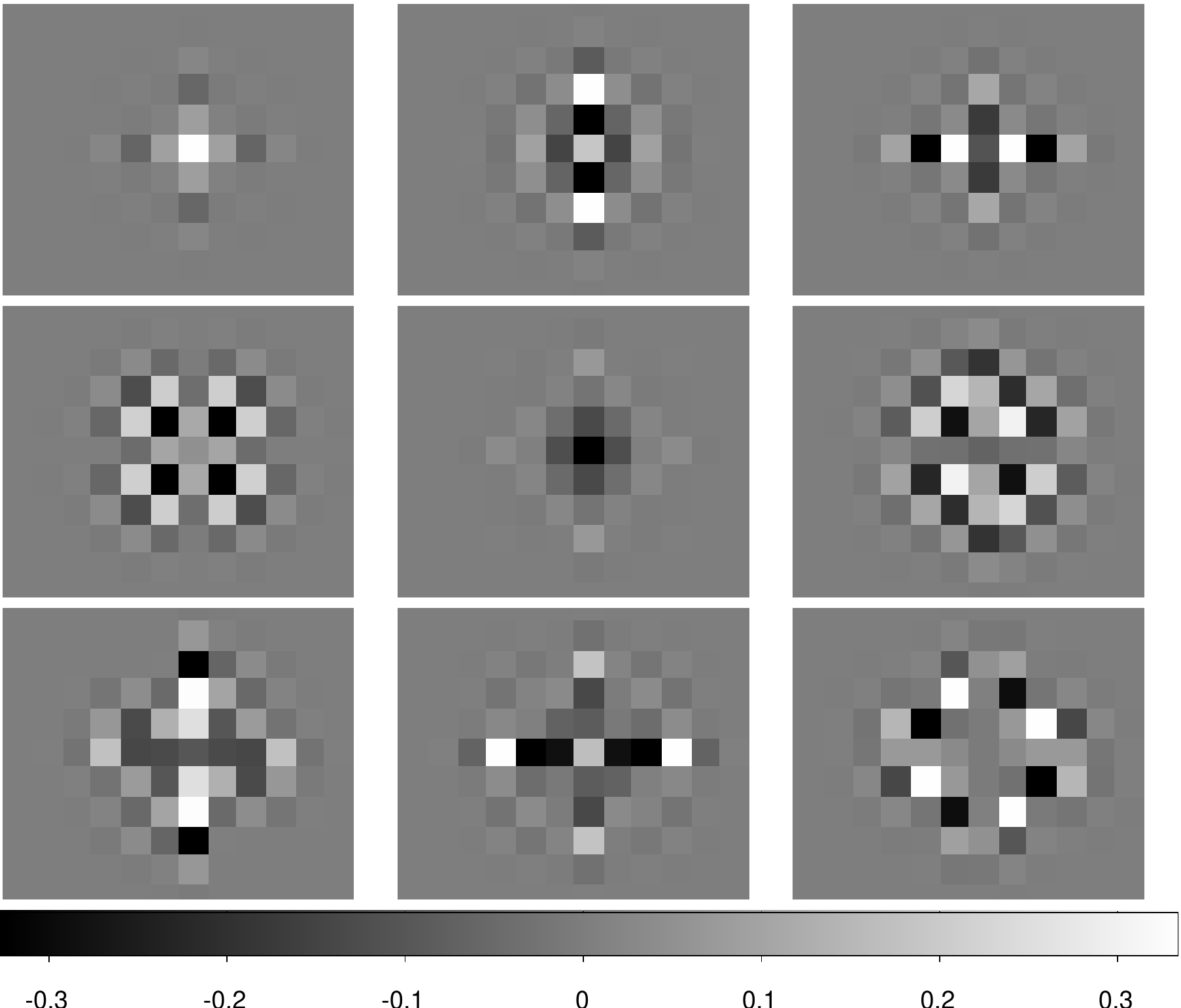}
	\caption{The upper left panel shows the average of noise correlation function on one coadd exposure. The rest of the panels show the first eight PCs of noise correlation functions on the coadd exposure.} \label{fig_noiCorPCs}
\end{figure}

Combining the HSC first year data with HSC-like image simulations, in this paper, we characterize several potential systematics in the FPFS shear estimation, which have not been discussed in \citet{Li18FPFS}.
One of them is the shear bias caused by PSF model residual. PSF model residual refers to the difference between the images of the true PSFs and the images of the PSF models reconstructed by the HSC pipeline \citep{HSC1-pipeline}. Unlike traditional methods such as KSB \citep{KSB95} and re-Gaussianization \citep{reGaussianization} that use moments of PSF images to correct smears of PSF, the FPFS algorithm deconvolves the PSF in Fourier space \citep[see][for similar treatment on PSF correction]{Z08,BFD14,metacal2} and therefore may have different dependence on the PSF model residual. Even though \citet{HSC1-catalog} have shown that the residuals of shapes and sizes on the coadd images of the HSC first year data meet the HSC first year science requirement\footnote{For the HSC first year science, the amplitude of the multiplicative bias is require to be smaller than $1.7 \%$. The amplitude of the correlation of additive bias is required to be smaller than $\xi^{\gamma}_{+}(\theta)/25$, where $\xi^{\gamma}_{+}$ is the shear-shear correlation.} \citep{HSC1-catalog}, the influence of the PSF model residual on the process of PSF deconvolution in Fourier space has not been directly quantified. To quantify the systematic error caused by the PSF residual to the FPFS shear estimator, we use the star images in HSC survey as input PSFs to convolve with the simulated galaxy images but use the corresponding PSF models at the positions of the stars for deconvolution in the procedure of the shape measurement \citep{LuPSF17}.
\begin{figure}
	\centering
	\textbf{Before Noise Subtraction}\par\medskip
	\includegraphics[width=.45\textwidth]{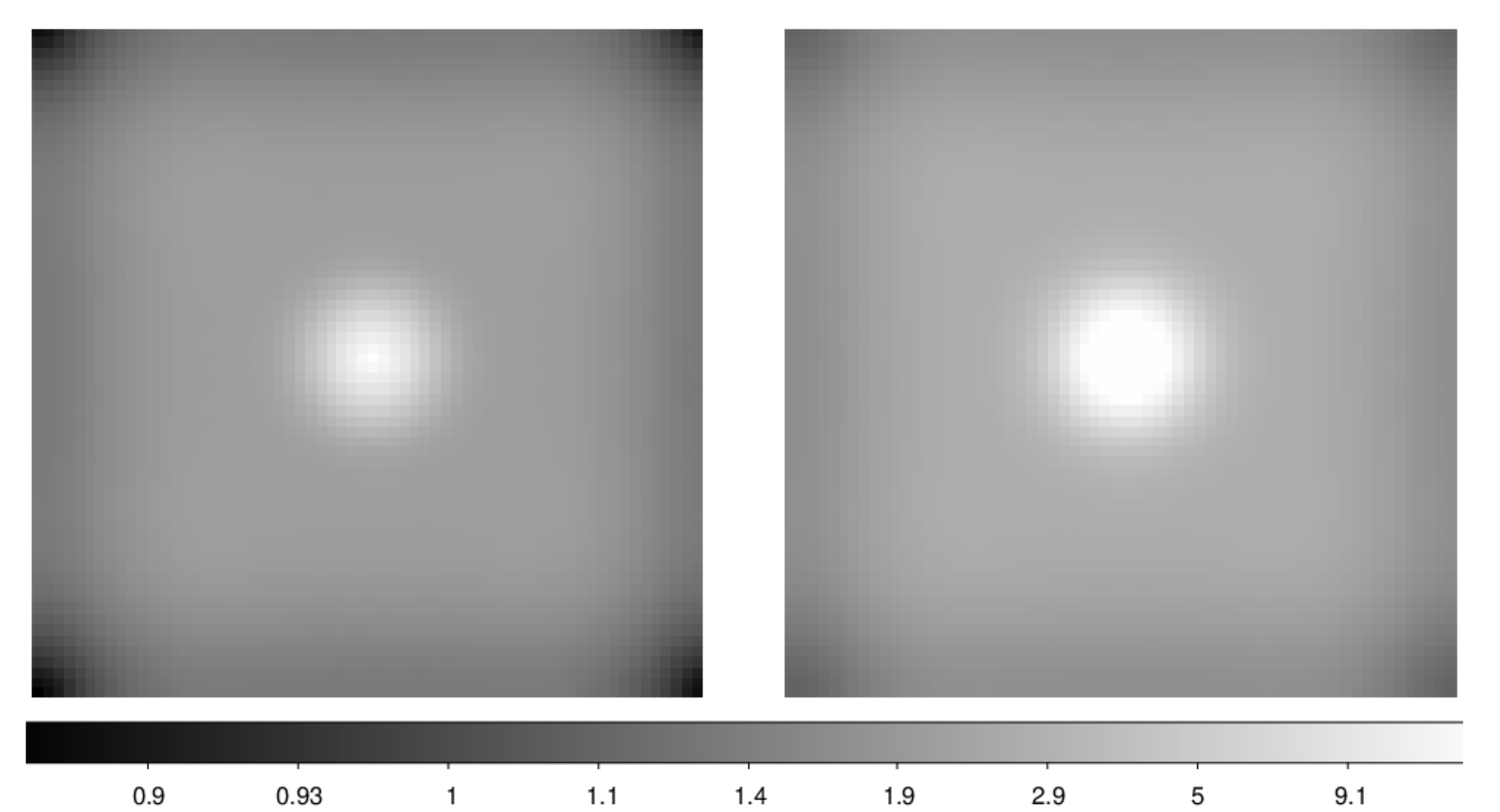}
	\textbf{After Noise Subtraction}\par\medskip
	\includegraphics[width=.45\textwidth]{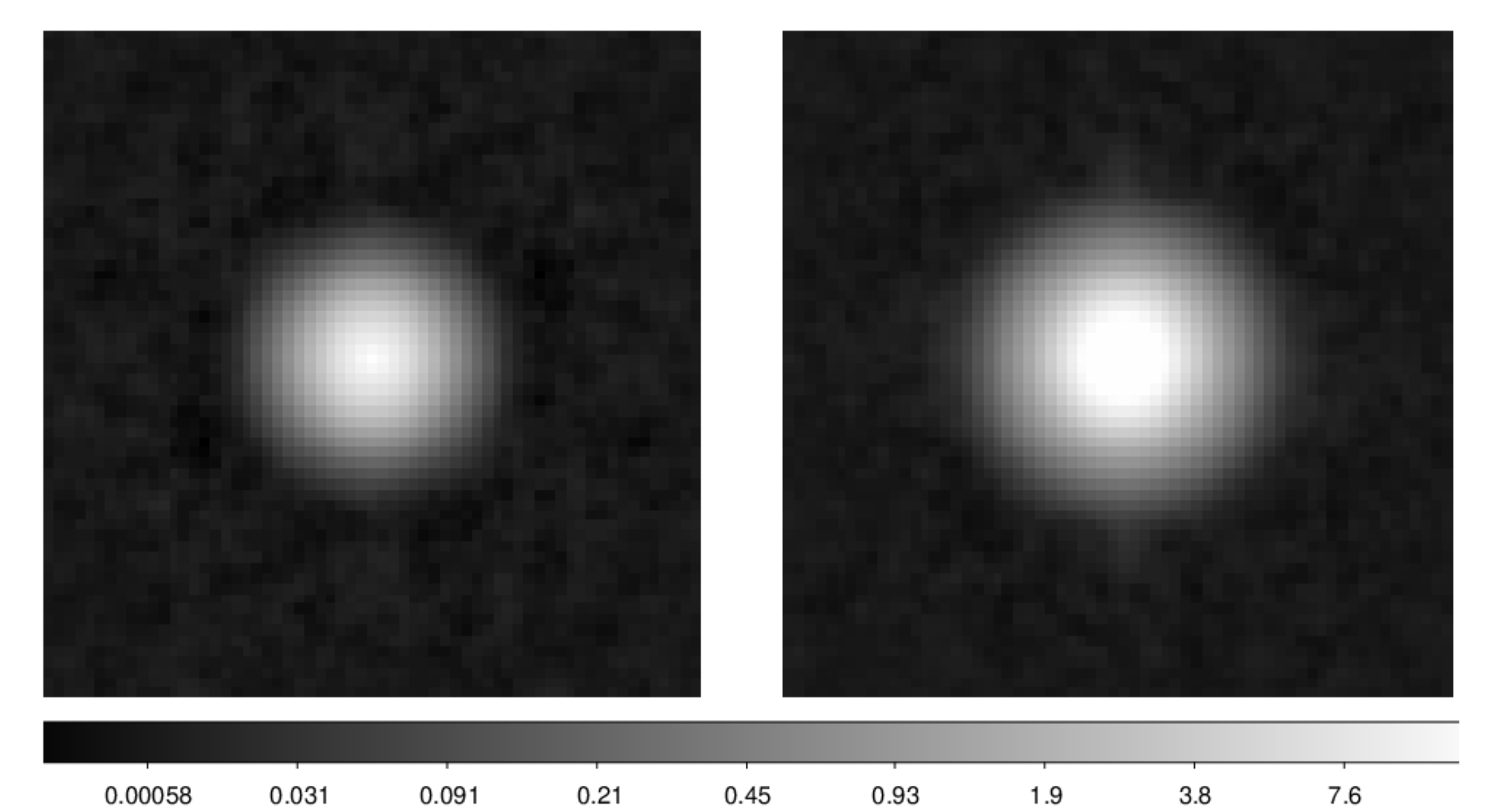}
	\caption{The stacked Fourier power functions of faint galaxies. The upper panels show the image before subtraction of Fourier power function of noise. The lower panels show the image after subtraction of Fourier power function of noise. The left panels show the stacked Fourier power function for galaxies with $\rm{S/N}<5$, where S/N is measured with the CModel algorithm \citep{HSC1-pipeline} . The right panels show the stacked Fourier power function for galaxies with $5\leq \rm{S/N} < 10$.} \label{fig_galMinFouPow}
\end{figure}
Another potential systematic is the calibration residual which refers to the bias caused by the difference between the simulated data used to calibrate the shear estimation and the observed data to which the calibrated shear estimator is applied to. In order to quantify the calibration residual, we apply the shear estimator calibrated by the default image simulation to simulations with different galaxy properties. We show that none of these bias exceeds the HSC first year science requirement.

\begin{figure*}
	\centering
	\includegraphics[width=.9\textwidth]{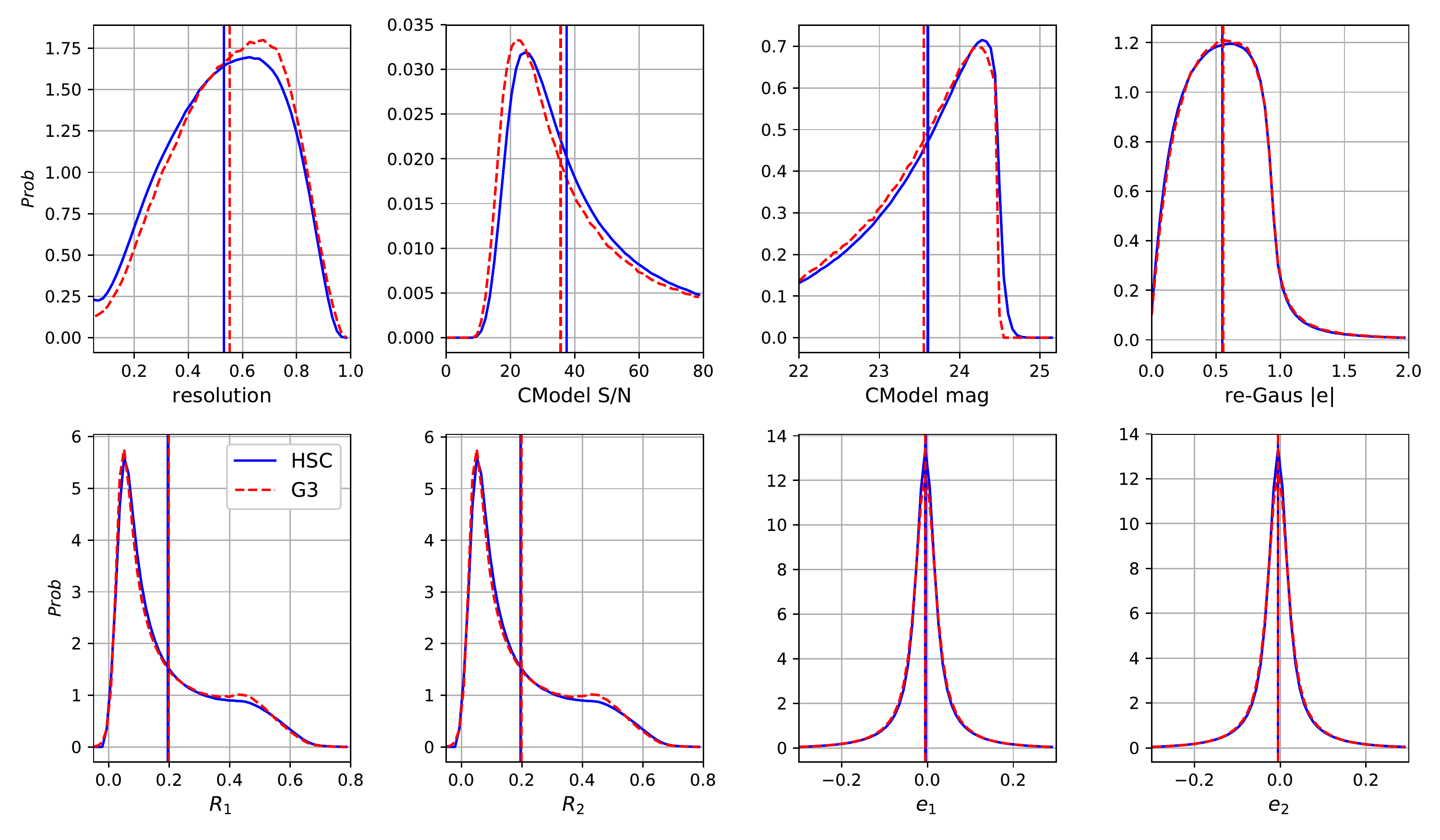}
	\caption{The histograms of eight observables including the re-Gaussianization resolution, the CModel S/N, the CModel magnitude, the re-Gaussianization distortion, two components of FPFS ellitpticity and two components of FPFS response. The solid lines show the histograms measured from HSC first year data and the dashed lines show the results from sample 4 of the HSC-like GREAT3 image simulations \citep{HSC1-GREAT3Sim}. The vertical lines show the average value of the corresponding observables.  All of the selection criteria shown in Table \ref{tab:source_selection} are applied to the HSC data. Only the $i$-band selection criteria in Table \ref{tab:source_selection} are applied to the simulation since we do not have multi-bands simulations. } \label{fig_EightHis}
\end{figure*}

Several internal null tests suggested by \citet{HSC1-catalog} are conducted on the FPFS shape catalog to demonstrate that the systematics on the FPFS shear estimator are below the level of the HSC first year science requirement \citep{HSC1-catalog}. We also compare the null test results of the FPFS shape catalog with those of the re-Gaussianization shape catalog \citep{HSC1-catalog} to check the consistency between these two catalogs.

We proceed with two major applications of the FPFS shape catalog, namely galaxy-galaxy lensing and mass map reconstruction. Firstly, we measure the galaxy-galaxy lensing signal for different lens catalogs using the FPFS shape catalog.
To further check the consistency between the FPFS shape catalog and the re-Gaussianization shape catalog, we compare our measurements with those of re-Gaussianization catalog. Subsequently, we apply our shape catalog to mass map reconstruction using the \citet{massMap-KS1993} method.

This paper is organized as follows. Section 2 describes the FPFS shape catalog based on the images of the HSC first year $i$-band Wide layer. Section 3 conducts several external systematic tests to quantify the bias caused by the PSF residual and the calibration residual. Section 4 performs internal null tests and compares the results with those of the re-Gaussianization shape catalog. Section 5 applies FPFS catalog to galaxy-galaxy lensing and compare the results with those of the re-Gaussianization shape catalog. Section 6 constructs mass maps with the FPFS shape catalog. Section 7 gives a conclusion. Throughout the paper, we adopt $\Omega_M=0.279$, $\Omega_b=0.046$, $\Omega_\Lambda=0.721$, $b=0.7$, $n_s=0.97$, and $\sigma_8=0.82$ \citep{WMAP9th}.

\section{the FPFS shape Catalog}
\label{sec:shearCatalog}

\subsection{Shear Estimator on Isolated galaxies}
\label{subsec:shearEstimator}
First we focus on the shape measurement of isolated galaxies. The systematic bias caused by blending is modeled and calibrated with HSC-like image simulations \citep{Li18FPFS}, which will be discussed in Section \ref{subsec:calib}.
\subsubsection{Fourier Power Function}
In order to reduce the uncertainty caused by pixel noise, FPFS algorithm defines the boundary of each galaxy using a circular top-hap aperture around the centroid of the galaxy \citep{Li17Auto} and sets the pixels outside the boundary to zero. The ratio between the aperture radius and the galaxy's half-light radius is set to a constant and denoted as $\alpha$. The half-light radius of each galaxy is calculated from the second order adaptive moment matrix measured by the re-Gaussianization algorithm \citep{reGaussianization} and the centroid is set to the center of the footprint of deblended galaxy. The Fourier power function of the galaxy is calculated as
\begin{equation} \label{eq:FourPowDef}
\begin{split}
\tilde{f}_o(\vec{k})&=\int f_o(\vec{x}_o) e^{-i\vec{k} \cdot \vec{x}_o} d^2x_o,\\
\tilde{F}_o(\vec{k})&=|\tilde{f}_o(\vec{k})|^2.
\end{split}
\end{equation}
\begin{figure*}
	\centering
	\includegraphics[width=.95\textwidth]{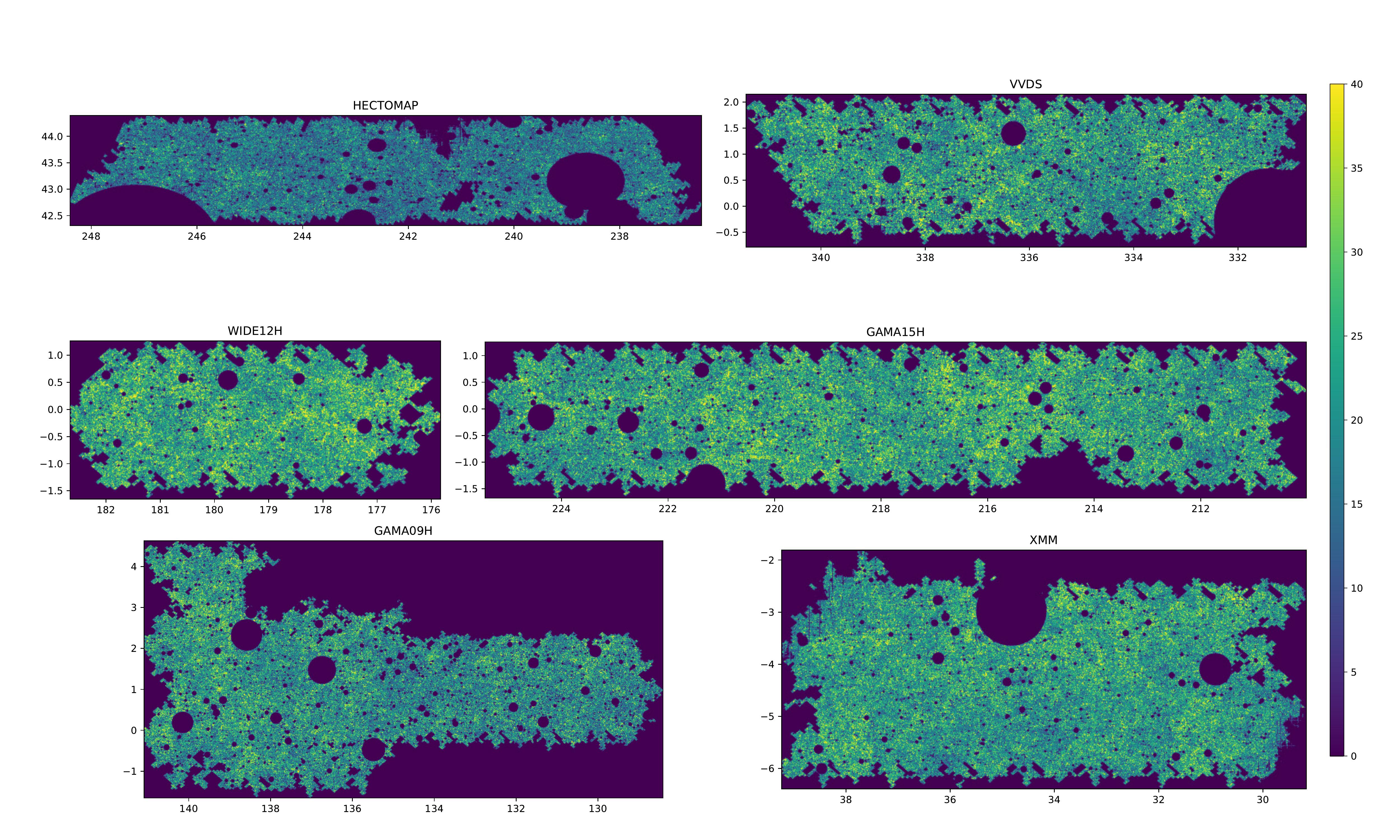}
	\caption{The number density maps of the FPFS galaxy shape catalog for six fields of the HSC first year data. The FPFS shape catalog contains more than $1.45 \times 10^7$ galaxies and the average number density is $28.6 ~ \rm{deg}^{-2}$.} \label{fig_nuMapNum}
\end{figure*}

The Fourier power function defined in equation (\ref{eq:FourPowDef}) is contaminated by the Fourier power function of noise.
The Fourier power function of noise depends on the correlation function of noise which is the weighted inverse Fourier transform of the noise Fourier power function \citep{Li18FPFS}.
Although noise on CCD images (single exposures) does not correlate across pixels \citep{Z15}, noise on coadd exposures correlates across pixels since an ad hoc warping kernel is used to convolve CCD images before re-pixelazation on a common coordinate in the coadding procedure.
Noise correlations on coadd exposures are mainly determined by shapes of the warping kernels.

For the HSC first year data, the HSC pipeline uses a third order Lanczos kernel to warp the CCD images, projects the warped images onto the common coordinates, and combines the projected images to generate coadd exposures \citep{HSC1-pipeline}.
The shapes of the projected warping kernels vary across the common coordinate since the projections of the input CCDs vary.
Noise correlation functions depend on the projected warping kernels, therefore they can be different for galaxies at different positions on coadd images.

In this paper we use Principal Component Analysis (PCA) to capture the variation of noise correlation functions on coadd exposures.
For each coadd exposure, we randomly sample one hundred positions and use the projected warping kernels at these positions to derive a group of noise correlation functions.
PCA is performed on such group of noise correlation functions. The standard deviation of these noise correlations as function of the rank of the principal components (PCs) is shown in Figure \ref{fig_noiCorPCsVar}.
The average correlation function and the first nine PCs are chosen as the basis vectors for noise correlation functions and will be used to model the power functions of noise, following \citet[][Appendix B]{Li18FPFS}, for galaxies at different positions on the coadd images.
As an example, we show the average noise correlation function and the first eight PCs for one of the exposures in Figure \ref{fig_noiCorPCs}.

Given that the HSC pipeline replaces neighboring sources with uncorrelated Gaussian noise before the shape measurement \citep{HSC1-pipeline}, the Dirac delta function is also added to the basis vectors of noise correlations with the intent to capture the influence of such neighboring source replacement to the correlation function of noise.

Using these basis vectors of noise Fourier power function, we fit the Fourier power function of each galaxy at large wave numbers, where the Fourier power function of the PSF decays to $10^{-4}$ of its maximum. Signals at such scale are mainly dominated by the power function of noise since the galaxy signal is filtered by the PSF and reduced to $10^{-4}$ of its original value. We denote the fitted model of noise Fourier power function for each galaxy as $\tilde{F}_n(\vec{k})$ and subtract it from the galaxy Fourier power function
\begin{equation}
\tilde{F}_r(\vec{k})=\tilde{F}_o(\vec{k})-\tilde{F}_n(\vec{k}).
\end{equation}

The upper panels of Figure \ref{fig_galMinFouPow} show the stacked Fourier power functions of faint galaxies before subtracting the noise Fourier power function. These images are significantly contaminated by the Fourier power function of noise, especially at large wave numbers. The lower panels of Figure \ref{fig_galMinFouPow} show the stacked Fourier power functions of faint galaxies after subtracting the noise Fourier power function. From the lower panels of Figure \ref{fig_galMinFouPow}, we conclude that our algorithm for for subtracting the noise Fourier power function works well on average at least at large wave numbers. We cannot directly see the performance at small wave numbers since the signals there are dominated by galaxies.

Since the residual of noise power function mainly influence shape measurement on faint galaxies, to mitigate the potential bias caused by imperfect noise power subtraction, conservative magnitude and S/N cuts are applied to our galaxy sample (see Section \ref{subsec:selection}). Also, conservative weighting scheme (see Section \ref{subsubsec:Shapelets}) is applied to our shape catalog. The next generation of HSC image simulations will includes different noise correlations and we will quantify the performance of our noise power function subtraction with image simulations.
\begin{figure*}
	\centering
	\includegraphics[width=.95\textwidth]{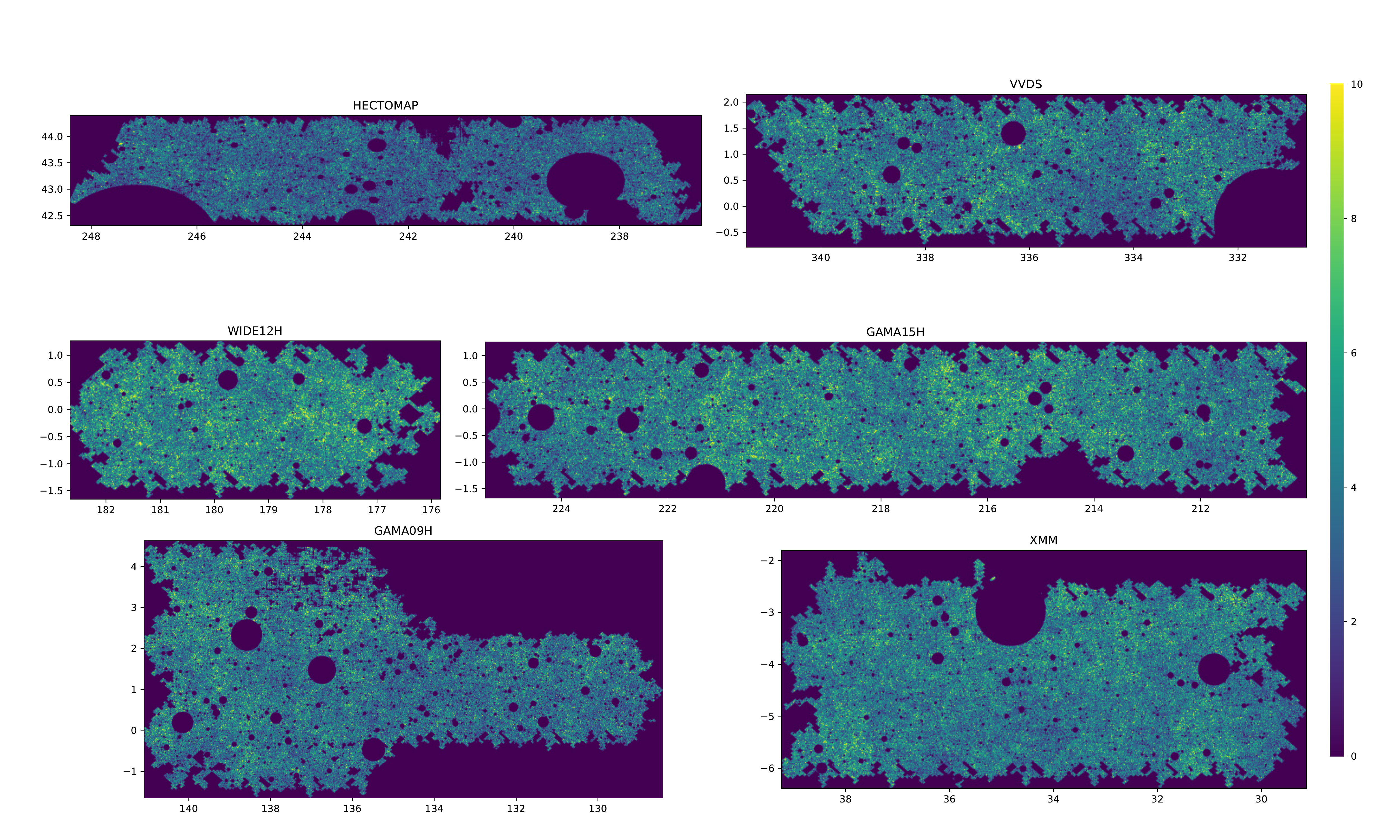}
	\caption{The density maps of the FPFS response ($R$) for six fields of the HSC first year data.} \label{fig_nuMapRes}
\end{figure*}

\subsubsection{Shapelets}
\label{subsubsec:Shapelets}
The PSF Fourier power function, denoted as $\tilde{G}(\vec{k})$, is subsequently deconvolved from the observed galaxy's Fourier power function to remove the influence of PSF
\begin{equation}\label{PSF deconvolution_fourier}
\tilde{F}(\vec{k})=\frac{\tilde{F}_r(\vec{k})}{\tilde{G}(\vec{k})}.
\end{equation}

Next, the deconvolved galaxy Fourier power function is projected onto the polar Shapelet basis vectors \citep{polar_Shapelets} as
\begin{align}\label{Shapelets_decompose}
M_{nm}=\int \chi_{nm}^{*} \tilde{F}(\rho,\phi) \rho d\rho d\phi.
\end{align}
The polar Shapelet basis vectors are defined as
\begin{align*}
\chi_{nm}(\rho,\phi)&=\frac{(-1)^{(n-|m|)/2}}{\sigma^{|m|+1}}\left\lbrace \frac{[(n-|m|)/2]!}{\pi[(n+|m|)/2]!}\right\rbrace^\frac{1}{2}\\
&\times \rho^{|m|}L^{|m|}_{\frac{n-|m|}{2}}\left(\frac{r^2}{\sigma^2}\right)e^{-\rho^2/2\sigma^2} e^{-im\phi},
\end{align*}
where $L^{p}_{q}$ are the Laguerre Polynomials, $n$ is the radial number and $m$ is the spin number, and $\sigma$ determines the scale of Shapelet functions. We denote the ratio between $\sigma$ and the scale radius of PSF Fourier power function ($r_{\text{pp}}$) as \citep{Li18FPFS}
\begin{equation}
\beta=\frac{\sigma}{r_{\text{pp}}}.
\end{equation}
Note that $\alpha$ determines the effective scale in configuration space whereas $\beta$ determines the effective scale in Fourier space. \citet{Li18FPFS} proposed to set $\alpha=4$, $\beta=0.85$ and showed that the systematic bias for such setup is below one percent of the shear signal.

Finally, using these shapelet modes, we define the dimensionless FPFS ellipticity and the corresponding shear response as
\begin{align}\label{ellipticity_define}
e_1=\frac{M_{22c}}{M_{00}+C},\qquad
e_2=\frac{M_{22s}}{M_{00}+C},\\
R_{1,2}=\frac{\sqrt{2}}{2}\frac{M_{00}-M_{40}}{M_{00}+C}+\sqrt{2}e_{1,2}^2,
\end{align}
$M_{nmc}$ and $M_{nms}$ are used to denote the real and imaginary part of $M_{nm}$ when $m>0$. The constant $C$ the weighting parameter which adjusts weight between galaxies with different luminosity and reduces noise bias \citep{Li18FPFS}. According to \citet{Li18FPFS}, we quantify the spread of $M_{00}$ with $\Delta$ which is the value of $M_{00}$ at which its histogram drops below $1/8$ of its maximum on the side of higher $M_{00}$ and normalize $C$ with $\Delta$ as $\nu = C/\Delta$. We conservatively set $\nu=4$ and the weight for galaxies with different S/N can be found from the left panel of Figure 7 in \citet{Li18FPFS}.

With the definition of average response $R= (R_1+R_2)/2$, the final shear estimator is
\begin{equation}
\gamma_{1,2} =-\frac{\left\langle e_{1,2} \right\rangle}{\left\langle R \right\rangle}.
\end{equation}
There exists a minus sign in the final shear estimator since the ellipticity are defined in Fourier space.

\subsection{Calibration for Blending Bias}
\label{subsec:calib}
\begin{table*}[!]
\centering
\begin{tabular}{cl}
\hline\hline
Selection & Explanation\\
\hline
\multicolumn{2}{c}{Cut on $i$-band properties}\\
\hline
ideblend\_nchild$==0$               & Do not contain child objects\\
iclassification\_extendedness $!=0$ & Mask out stars\\
idetect\_is\_primary $==$ True      & Identify unique detections only\\
ideblend\_skipped $==$ False        & Deblender skip the source object\\
bad\_pixel\_aperture $==$  False    & No bad pixel within the aperture \footnote{Bad pixels include `BAD', `SAT', `INTRP', `CR', `NO\_DATA', `SUSPECT', `UNMASKEDNAN', `CROSSTALK', `NOT\_DEBLENDED', `DETECTED NEGATIVE' pixels.}\\
iflux\_cmodel/iflux\_cmodel\_err$\geq 10$  &Galaxy has high enough S/N in i band\\
imag\_cmodel$- a_i < 24.5$ & Cut on Magnitude\\
$|e_{\rm psf}|<0.2$,           & Cut on PSF shape  \\
$0.39''<\text{FWHM}<0.79''$& Cut on PSF size   \\
$s>0.067$                  & Cut on FPFS flux ratio\\
$|R|<5$                    & Cut on FPFS response\\
iblendedness\_abs\_flux $< 10^{-0.375}$ &spurious detection and heavy blending\\
\hline
\multicolumn{2}{c}{Cut on multi-bands properties}\\
\hline
FDFC $==$ True & Select sources within the FDFC\footnote{FDFC refers the Full Depth Full Color cut defined in \citet{HSC1-catalog}} region\\
\multicolumn{2}{c}{Require that at least two of the following four cuts be passed}\\
gflux\_cmodel/gflux\_cmodel\_err$\geq 5$  &Galaxy has high enough S/N in g band\\
rflux\_cmodel/gflux\_cmodel\_err$\geq 5$  &Galaxy has high enough S/N in r band\\
zflux\_cmodel/gflux\_cmodel\_err$\geq 5$  &Galaxy has high enough S/N in z band\\
yflux\_cmodel/gflux\_cmodel\_err$\geq 5$  &Galaxy has high enough S/N in y band\\
\hline\hline
\end{tabular}
\caption{Selection criteria of source galaxies.}\label{tab:source_selection}
\end{table*}
As observations go deeper, the density of detected sources increases.
Due to the limited resolution of ground-based telescopes, there are high possibilities that multiple sources are blended within one detected footprint for deep surveys
such as HSC \citep{HSC1-pipeline}. If such blending happens, we deblend the footprint and measure the shear from each isolated source separately.
Since these blended sources can be located at different redshifts, sources can be distorted by different shear signals even though they are close to each other on the
transverse two-dimensional sky plane.

The HSC pipeline \citep{HSC1-pipeline} uses the SDSS deblender \citep{SDSSpipe} to isolate the sources within one footprint if multiple number of sources are detected
within the footprint. The HSC pipeline replaces all of the neighboring sources with uncorrelated Gaussian noise. Finally, we apply the FPFS shear estimator to the isolated
galaxy images.

\citet{Li18FPFS} uses HSC-like galaxy image simulations \citep{HSC1-GREAT3Sim} to estimate the bias caused by blending. The image simulation is described as follows.
Postage stamps from the COSMOS Hubble Space Telescope (HST) survey ACS field \citep{HST-ACSpipe,HST-shapeCatalog-Alexie2007} are selected according to the positions
of galaxies detected from the HSC Wide-depth coadds overlapping with the COSMOS region \citep{HSC1-data}. Subsequently, these image stamps are deconvolved, distorted
by known shear $\gamma_{1,2}$, convolved with HSC-like PSFs and contaminated with HSC-like noise to mimic HSC images.  The simulated HSC galaxies are separated into
$800$ subfields. Different subfields have different HSC-like PSFs, distorted by different shear, and contaminated with different HSC-like noise \citep{HSC1-GREAT3Sim}.
{\oColorS{
The input postage-stamps of the HSC-like simulation \citep{HSC1-GREAT3Sim} keep the information on the COSMOS-HST images without deblending and masking out the neighboring
sources. The inclusion of neighboring sources is critical to reproducing the observed distributions of galaxy sizes and magnitudes.
}}

{\bColor{
However, we note that there are two limitations of the simulation.
The first limitation is that the background information beyond the scale of the postage stamp ($64 \times 0.168 ~\rm{arcsec}$)
is neglected in our simulation. Such global background could influence the galaxy detection and the shape measurement of faint galaxies. The influence should differs among
different background subtraction algorithms.

The second limitation is that even though the HSC-like simulation keeps all of the local information of neighboring sources recorded on the COSMOS-HST exposures, due to
the limited depth of the COSMOS-HST survey, galaxies fainter than $25.2$ can not be distinguished from noise. Therefore the input galaxy sample is not complete at the
faint end of magnitude $~$ 25.2.
\citet{metaDet2019} finds a selection bias caused by the shear dependent galaxy detection. The shear dependent galaxy detection refers to the fact that the possibility of
distinguishing two neighboring sources correlates with the shear distortion.
Such shear dependent detection leads to a few percent multiplicative bias. According to \citet{metaDet2019}, it is important to include the information of neighboring sources
in the simulation to calibrate the selection bias caused by shear dependent galaxy detection.
The calibration derived from our simulation contains the shear dependent selection bias from neighboring sources which are brighter than $25.2$ but the shear dependent selection
bias from neighboring sources which are fainter than $25.2$ may not be fully included in the calibration.

{\oColorS{
This work applies a $24.5$ CModel magnitude cut as suggested by \citet{HSC1-GREAT3Sim} to mitigate the potential biases caused by the limitations of the simulation.
As shown in Figure \ref{fig_EightHis}, the simulation reproduces the observed histograms of galaxy sizes, magnitude and also the FPFS shapes, etc. These limitations
do not cause significant systematic biases on the measurements of these observables. Therefore, we expect that these limitations do not significantly bias the shear
calibration.
Further discussions on the influences of global background and shear correlated detection bias from galaxies fainter than $25.2$ are left to the future works.
}}
}}

The deblender, neighboring source replacer and shear estimator are successively run on the simulated galaxies to measure the shear and the measured shear is denoted as $\gamma^M_{1,2}$. We assume a linear relation between the measured shears and the input shears based on the fact that the amplitudes of shear signals are only a few percent for weak lensing
\begin{equation}\label{eq:shear_calibration}
\gamma^{M}_{1,2}=(1+m_{1,2}) \gamma_{1,2}+c_{1,2},
\end{equation}
where $m_{1,2}$ are multiplicative biases for two shear components and $c_{1,2}$ are additive biases for two shear components. We use $m$ to denote the mean of multiplicative biases on two shear components, where $m= (m_1+m_2)/2$. The values of $m$ and $c_{1,2}$ depend on galaxy properties, which is modeled with image simulations.

Note that \citet{HSC1-GREAT3Sim} further model the additive bias as $c_{1,2} = a_{1,2}e^{p}_{1,2}+c'_{1,2}$, considering the correlation between additive bias and PSF ellipticity, where $e^{p}_{1,2}$ are two components of the PSF ellipticity, $a_{1,2}$ are two components of the fractional additive bias, $c'_{1,2}$ are two components of the remaining additive bias. As demonstrated in Appendix \ref{Appendix:corCeP}, the amplitude of $a_{1,2}$ is below $0.5\%$ for the FPFS shape estimator, which is below the HSC first year science requirement. Therefore, we neglect $a_{1,2}$ in this paper.

\citet{Li18FPFS} reported that, for HSC-like galaxy image simulations \citep{HSC1-GREAT3Sim}, a multiplicative bias of $-5.8 \pm 0.4 \%$ on average is found on the aforementioned pipeline and no additive bias on average beyond the statistical uncertainty was found due to the blending effect.
{\bColor{There exists two causes of such multiplicative bias.
The first is that the SDSS deblender does not accurately recover the true galaxy shapes from blended footprints and we refer it to as deblending bias. The other is that, as shown in \citet{metaDet2019}, shear-dependent detection in the presence of blending leads to a few percents multiplicative bias at HSC-like object density.
Here we do not separate the deblending bias from the shear-dependent detection bias and leave it to our future work.}}
\begin{table}
\centering
\begin{tabular}{cl}
\hline\hline
Column name & Explanation\\
\hline
$\rm{fpfs}\_e_{1,2}$                       & FPFS ellipticity\\
$\rm{fpfs}\_\rm{RA}$                       & Average response\\
$\rm{fpfs}\_\rm{flux}$                     & FPFS flux ratio\\
$\rm{fpfs}\_\rm{m}$                        & FPFS multiplicative bias\\
\hline\hline
\end{tabular}
\caption{Columns for the FPFS shapes.}\label{tab:FPFS_shapes}
\end{table}

The modeling and calibration of blending bias is conducted as follows. Galaxies are first divided into different bins according to FPFS flux ratio\footnote{We termed $s$ FPFS flux in \citet{Li18FPFS} but it is incorrect since $s$ is dimensionless. Therefore, we rename $s$ as FPFS flux ratio.} ($s$)
\begin{equation}\label{eq:FPFS_flux define}
s=\frac{M_{00}}{M_{00}+C}.
\end{equation}
Next, shear is estimated independently for galaxies in each bin. By fitting the linear relation between the input shear ($\gamma$) and the estimated shear ($\gamma^M$) with equation (\ref{eq:shear_calibration}), the average multiplicative bias is measured for each FPFS flux ratio bin. Finally, multiplicative bias on individual galaxy is modeled as a third order polynomial function of FPFS flux ratio
\begin{equation}
\hat{m}=a_3 s^3+a_2 s^2+a_1 s^1+a_0,
\end{equation}
where $a_i (i=0,1,2,3)$ are determined by fitting FPFS flux ratio to the average multiplicative bias in each bin \citep{Li18FPFS}. The calibrated shear estimator is
\begin{equation}
\gamma_{1,2}=-\frac{\left\langle e_{1,2} \right\rangle}{\left\langle (1+\hat{m})R \right\rangle}.
\end{equation}

\begin{figure*}
\centering
\includegraphics[width=.9\textwidth]{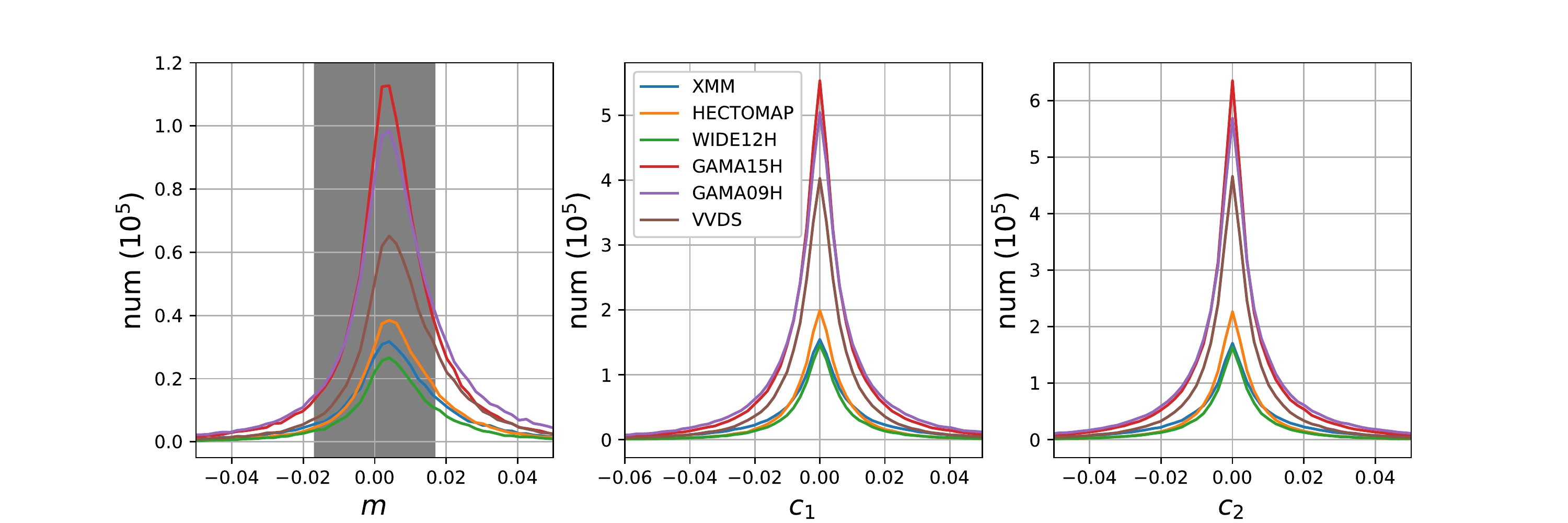}
\caption{The histograms of multiplicative bias and two components of additive bias caused by the PSF model residual for six fields of the HSC first year data. The grey area shows the HSC first year science requirement.} \label{fig_psfResMcHis}
\end{figure*}

\begin{table*}
\centering
\begin{tabular}{ccccccc}
\hline \hline
field & $\langle m \rangle (10^{-2})$& $\sqrt{\langle m^2\rangle}(10^{-2})$&$\langle c_1 \rangle (10^{-4})$ & $\sqrt{\langle c_1^2\rangle}(10^{-4})$&  $\langle c_2 \rangle (10^{-4})$& $\sqrt{\langle c_2^2\rangle}(10^{-4})$  \\ \hline
XMM & $0.56$ & $4.78$ & $0.02$ & $180.65$ & $0.15$& $187.55$\\
GAMA09H & $0.73$ & $3.67$ & $-0.41$ & $171.58$ & $0.84$& $169.51$\\
GAMA15H & $0.51$ & $2.78$ & $-1.66$ & $137.95$ & $0.35$& $137.90$\\
HECTOMAP & $0.76$ & $2.80$ & $-0.72$ & $124.67$ & $0.96$& $132.93$\\
WIDE12H & $0.64$ & $2.61$ & $0.42$ & $123.79$ & $0.27$& $133.53$\\
VVDS & $0.81$ & $2.91$ & $-2.02$ & $136.52$ & $0.95$& $132.97$\\
 \hline
\end{tabular}
\caption{The averages and Root-Mean-Squares of the multiplicative biases and additive biases for six fields of the first year data of HSC survey. The averages of multiplicative biases are within the HSC first year science requirement, which is $1.7\%$.}\label{tab:psfMc}
\end{table*}

We model the multiplicative bias as a function of FPFS flux ratio and marginalize its dependence over other properties.
It is important to note that the multiplicative bias should also depend on properties other than FPFS flux ratio.
If the simulation has the same galaxy distribution as the observed galaxy sample over the marginalized properties, such calibration provides us with an unbiased shear estimation.
However, if this condition is not met, the average multiplicative bias in each FPFS flux ratio bin estimated from the simulation could be biased from that of the observed galaxy sample so that the calibration could be biased.
We refer to such bias as calibration residual.

To mitigate the calibration residual, it is necessary to ensure the galaxy distribution over marginalized properties are the same between the simulation and the observation.
Figure \ref{fig_EightHis} shows the galaxy distribution over eight different properties of HSC data and in our HSC-like simulation. We find that the distribution are indeed quite similar between simulations and observation. Moreover,
Section \ref{subsec:calibTest} quantifies the amplitude of the potential calibration residual caused by the differences between simulations and observed galaxy ensemble.

\begin{figure*}
\centering
\includegraphics[width=.9\textwidth]{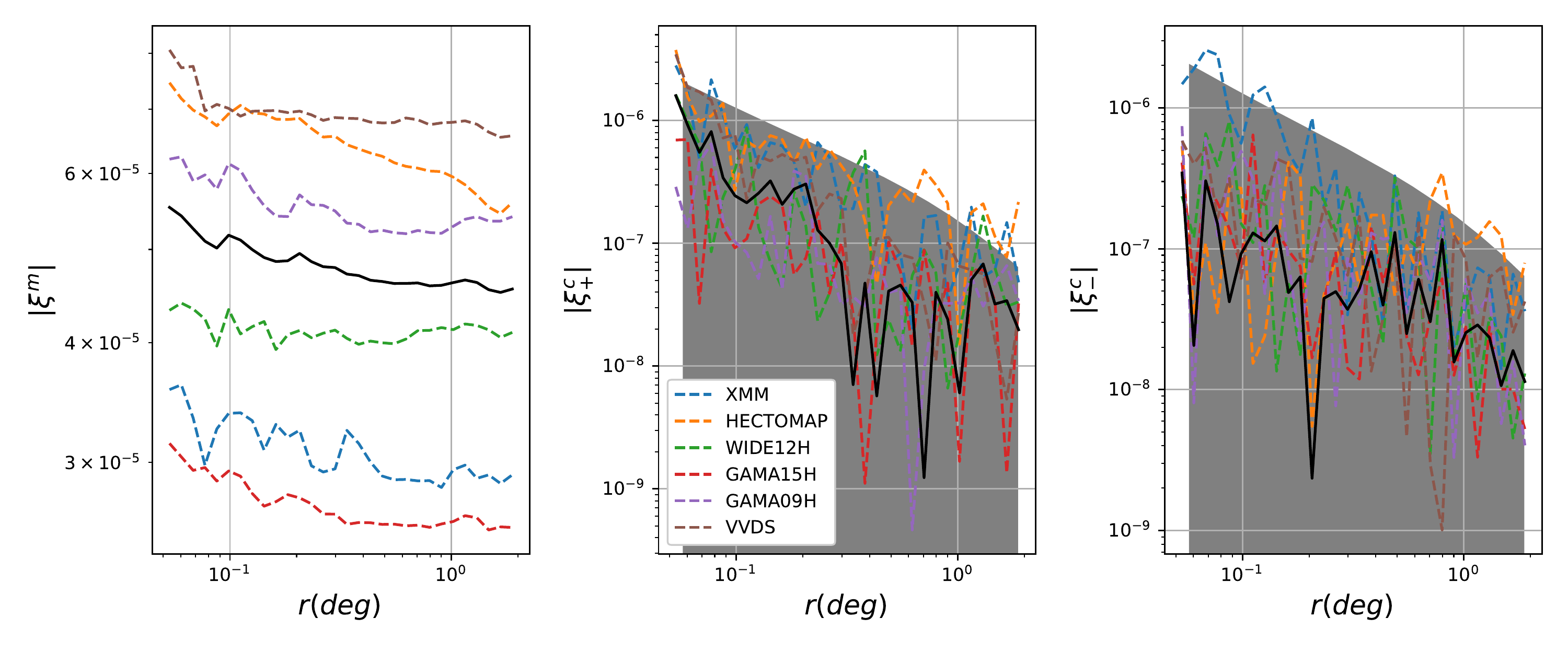}
\caption{The spatial correlations of multiplicative bias and additive bias caused by the PSF residual. The left panel shows the correlation function of multiplicative bias. The middle panel and the right panel show the two components of correlation function for additive bias. Colorful dashed lines show the absolute value of correlation functions measured from different field of the HSC first year survey. Black solid lines show the absolute value of average correlation functions. Grey areas show the fisrt year HSC science requirement.} \label{fig_psfResMcSpaCor}
\end{figure*}
\subsection{Galaxy Selections}
\label{subsec:selection}
Table \ref{tab:source_selection} summarizes the source selection criteria for the FPFS shape catalog. The number density of galaxies and the density of response after the selection are shown in Figure \ref{fig_nuMapNum} and Figure \ref{fig_nuMapRes}, respectively.  The FPFS shape catalog contains more than $1.45 \times 10^7$ galaxies and the average number density is $28.6 ~ \rm{deg}^{-2}$.

In comparison with the selection criteria for the re-Gaussianization shape catalog \citep[][Table 4]{HSC1-catalog}, we do not select galaxies on resolution as done in the re-Gaussianization catalog. Therefore, we do not need to include the selection bias caused by the cut on resolution. However, galaxies are selected according to the shape ($|e_{\rm psf}|$) and size ($\text{FWHM}$) of the PSFs. Since $|e_{\rm psf}|$ and $\text{FWHM}$ do not correlate with the direction of shear nor have preference on any direction, the intrinsic orientations of galaxies selected by these cuts are random. Therefore, such selections do not induce any selection bias.

Moreover, we mask out the galaxy if the HSC pipeline \citep{HSC1-pipeline} detected bad pixels inside the top-hat aperture. On the other hand, the re-Gaussianization shape catalog mask out only galaxies with bad pixels near the centroids of galaxies.

We add the same $i$-band magnitude cut and S/N cut as in the re-Gaussianization shape catalog. In addition, we add a cut on the FPFS flux ratio, where $s$ defined in equation (\ref{eq:FPFS_flux define}). We confirm that the selection bias caused by these selections is well below the HSC first year science requirement.
\subsection{Shape catalog}
\label{subsec:FPFS_catalog}
The FPFS catalog contains the FPFS shapes as well as all of the columns shown in Table 3 of \citet{HSC1-catalog}. The columns for FPFS shapes are listed in Table \ref{tab:FPFS_shapes}. Note that there are two main differences between the FPFS shapes and re-Gaussianization shapes. The first difference is that the FPFS shear estimator does not apply the lensing weight. As a result, it does not rely on external simulations for the weight bias calibration. The second difference is that the FPFS shear estimator does not need to calibrate additive biases, at least for the accuracy required by the HSC first year science, since since additive biases are quite small on average and do not correlate with the shape of PSF \citep{Li18FPFS}.

Section \ref{subsec:ESD} describes the procedure for using the FPFS shape catalog to measure the average excess surface densities (ESDs) in galaxy-galaxy lensing. Section \ref{subsec:kappaMap-KS} describes the procedure for using the FPFS shape catalog to reconstruct a mass map with the Kaiser-Squires method \citep{massMap-KS1993}. Although in this paper we do not provide cosmic shear measurement using the FPFS shape catalog, the procedure of using the FPFS shape catalog to measure cosmic shear is described in Appendix \ref{Appendix:cosmic shear}.

\section{External Tests}
\subsection{The PSF Residual}
\label{subsec:psfTest}
The HSC pipeline uses restructured version of PSFEx \citep{PSFEx11} to perform a polynomial fit of the PSF as a function of position on every CCD. The stars used to feed PSFEx are selected by a k-means clustering algorithm on each CCD \citep{HSC1-pipeline}. PSF models for galaxies on coadd exposures are constructed using all of the corresponding PSF models from the input CCDs at the same sky coordinates \citep{HSC1-pipeline}. We refer to the difference between the true PSF and the PSF model as PSF residual.

\citet{HSC1-pipeline} and \citet{HSC1-catalog} tested the PSF modeling on single visit level and coadd level, respectively, by comparing the size and shape between stars and interpolated PSF model at the positions of stars. They checked the probability distribution functions (PDFs) and the spatial correlations of size residual and ellipticity residual for the PSF model and conclude that these residuals meet the HSC first year science requirement \citep{HSC1-catalog}.

It is important to note that FPFS algorithm does not use moments of PSF models to remove the influence of PSFs to the shear estimator. Instead, it deconvolves PSFs in Fourier space, as done in several other recently developed shear estimators \citep[eg.][]{Z08,BFD14,metacal2}. In what follows, we conduct the systematic tests suggested by \citet{LuPSF17} to directly quantify the influence of the PSF residual on the FPFS shear estimation using star images in HSC survey and external galaxy image simulations.

The setups of the PSF tests are described as follows. We first select stars with $S/N$ greater than $500$, $i$-band magnitude less than $22.5$. We also require the stars to have the measurement of re-Gaussianization moments and to not contain bad pixels near the centroid. For each star, a group of noiseless modeled galaxies fitted to the $25.2$ magnitude limited COSMOS HST galaxy sample \footnote{\url{great3.jb.man.ac.uk/leaderboard/data/public/COSMOS_25.2_training_sample.tar.gz}} are subsequently simulated and distorted with input shears using Galsim which is an open-source software for galaxy image simulation \citep{Galsim}. To remove the shape noise, the galaxies are grouped in orthogonal pairs, whose intrinsic orientations is $90$ degree apart from each other. Then the galaxies are convolved with each star image. We use the PSF model reconstructed by the HSC pipeline at the corresponding position of the star to deconvolve the simulated galaxies and measure shear with the FPFS shear estimator. Finally, a linear fitting from the input shear to the measured shear with equation (\ref{eq:shear_calibration}) is performed to determine the multiplicative bias $(m_{1,2})$ and additive bias $(c_{1,2})$ for each star selected by the test. Note that we neglect the difference between two components of the multiplicative bias since it is quite small. The average of two components of multiplicative bias $m=(m_1+m_2)/2$ is used for our analysis.

\begin{figure*}
\centering
\includegraphics[width=.9\textwidth]{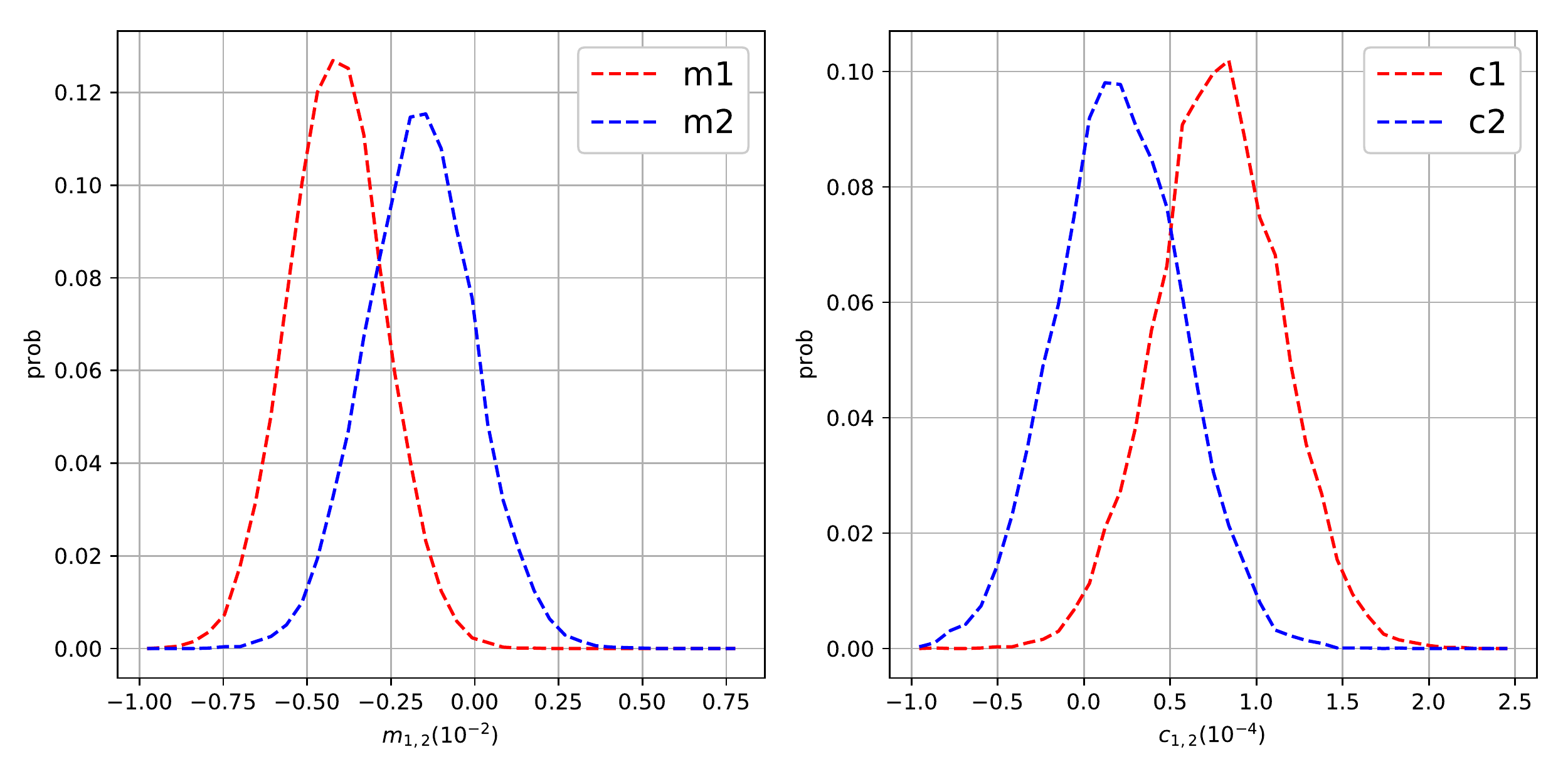}
\caption{The probability distributions of the calibration residual. The left panel shows the distributions of
residual multiplicative biases, wheres the right panel shows those of residual additive biases. } \label{fig_calibResiduals}
\end{figure*}

One reason to use bright stars with $S/N>500$ is to reduce the shear measurement bias caused by the noise on star images. Since the simulated galaxies are noiseless and isolated, this PSF test focuses on the shear measurement bias caused by the PSF residual. Another reason to use the bright star is to ensure the pureness of the star sample. The extendedness algorithm \citep{HSC1-pipeline} used to distinguish stars from galaxies performs well on the bright sources, whereas on the faint end there could be galaxies mixed into the star sample.

The histograms of multiplicative bias and additive bias caused by the PSF model residual for six different fields in the HSC first year data are shown in Figure \ref{fig_psfResMcHis}. The averages and root-mean-squares of the multiplicative bias and additive bias are laid out in Table \ref{tab:psfMc}. We find that the averages of the multiplicative bias is below $1.7\%$, which meet the HSC first year science requirement.

\begin{figure*}
\centering
\includegraphics[width=.9\textwidth]{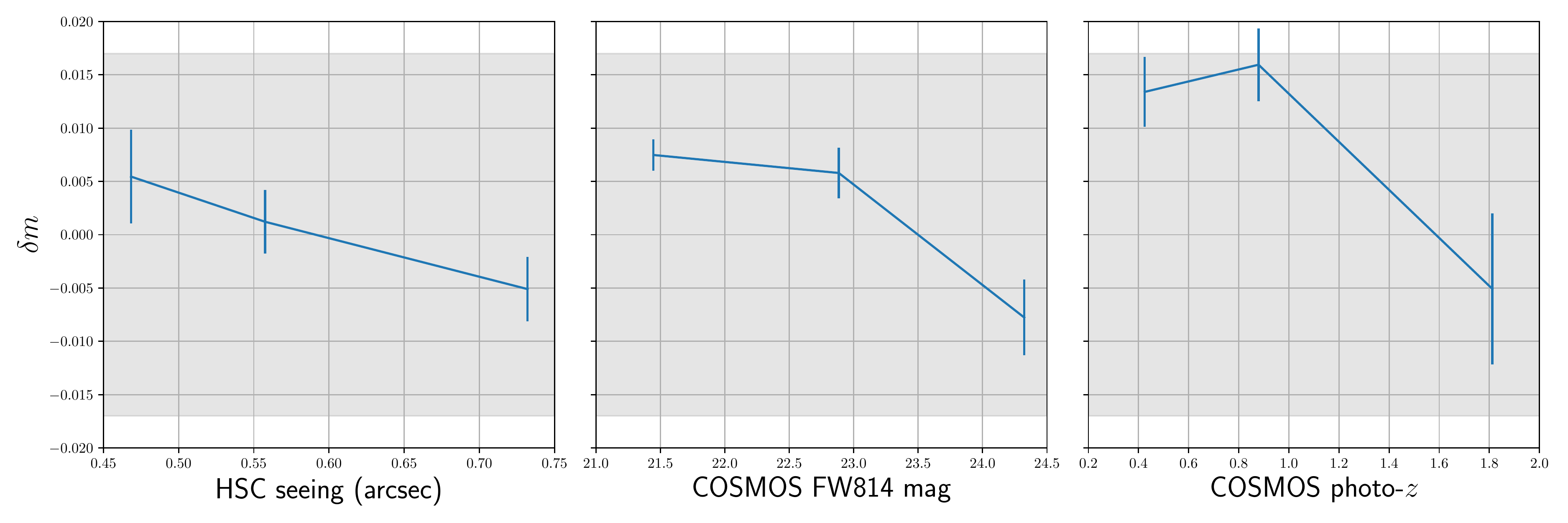}
\caption{Correlations of calibration residual with seeing, magnitude, and photo-$z$. The left panel shows the calibration residuals for galaxies in three seeing bins, the middle panel shows the calibration residuals for galaxies in three COSMOS F814W magnitude bins, and the right panel shows the calibration residuals for galaxies in three COSMOS photo-$z$ bins. The gray area denotes the first year HSC science requirement.} \label{fig_calibResiduals_mag_z_fwhm}
\end{figure*}

In addition to the PDFs of the biases, we also check the spatial correlation functions of the biases caused by the PSF residual. The two components of the measured shear correlation function ($\hat{\xi}^\gamma_\pm$) are influenced by the multiplicative bias and additive bias as
\begin{eqnarray}\label{eq:shearCor_psfres}
\hat{\xi}^\gamma_{\pm}&&=\left\langle \hat{\gamma}_t\hat{\gamma}_t\right\rangle \pm \left\langle \hat{\gamma}_\times \hat{\gamma}_\times\right\rangle\\\notag
&&=(1+2\left\langle m \right\rangle+\xi^m )\xi^\gamma_{\pm} + \xi^c_{\pm},
\end{eqnarray}
where $\hat{g}_t$ is the component along or perpendicular to the separation (tangential component), and $\hat{g}_\times$ is the component at 45 degree (cross component). $\xi^\gamma_{\pm}$ represent two components of the true shear-shear correlation function. The correlations of multiplicative bias and additive bias ($\xi^m$ and $\xi^c_{\pm}$) are defined as
\begin{eqnarray}
\xi^m && = \langle m m \rangle,\\
\xi^c_{\pm} &&= \langle c_t c_t \rangle \pm \langle c_{\times} c_{\times} \rangle.
\end{eqnarray}

The correlation functions of multiplicative bias and additive bias for six fields in the HSC first year survey are shown in Figure \ref{fig_psfResMcSpaCor}. \citet{HSC1-catalog} requires $|\xi^c_{\pm}|<\xi^{\gamma}_{+}/25$ to ensure the systematic bias is below the statistical uncertainty, where $\xi^{\gamma}_{+}$ is the expected shear-shear correlation function. Such requirement is shown as grey areas in Figure \ref{fig_psfResMcSpaCor}.
We conclude that the biases introduced by the PSF model residual are within the HSC first year science requirement.

\subsection{The Calibration Residual}
\label{subsec:calibTest}
\begin{figure*}
\centering
\begin{tabular}[b]{@{}p{0.45\textwidth}@{}}
\includegraphics[width=.45\textwidth]{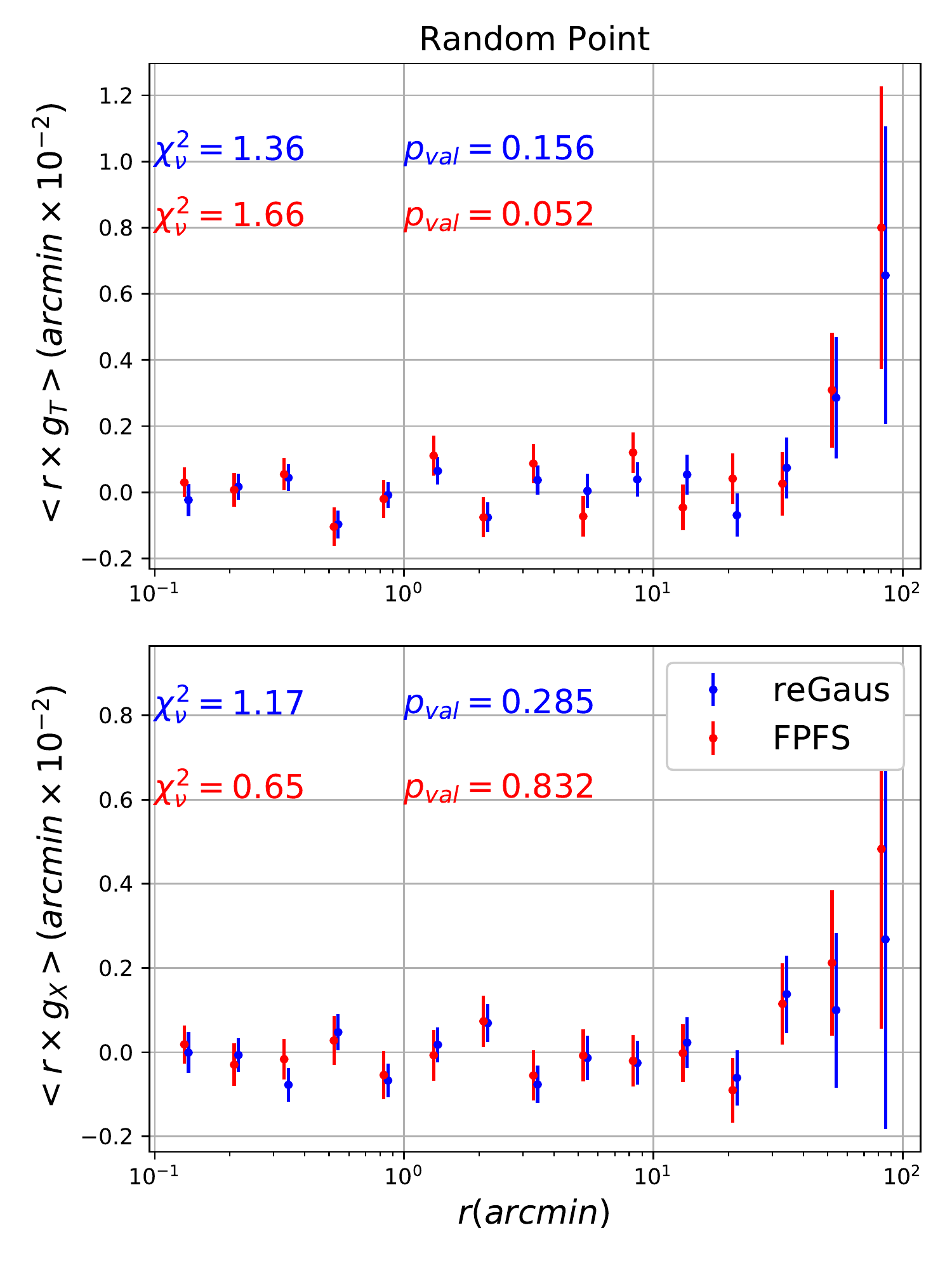}
\end{tabular}
\begin{tabular}[b]{@{}p{0.45\textwidth}@{}}
\includegraphics[width=.45\textwidth]{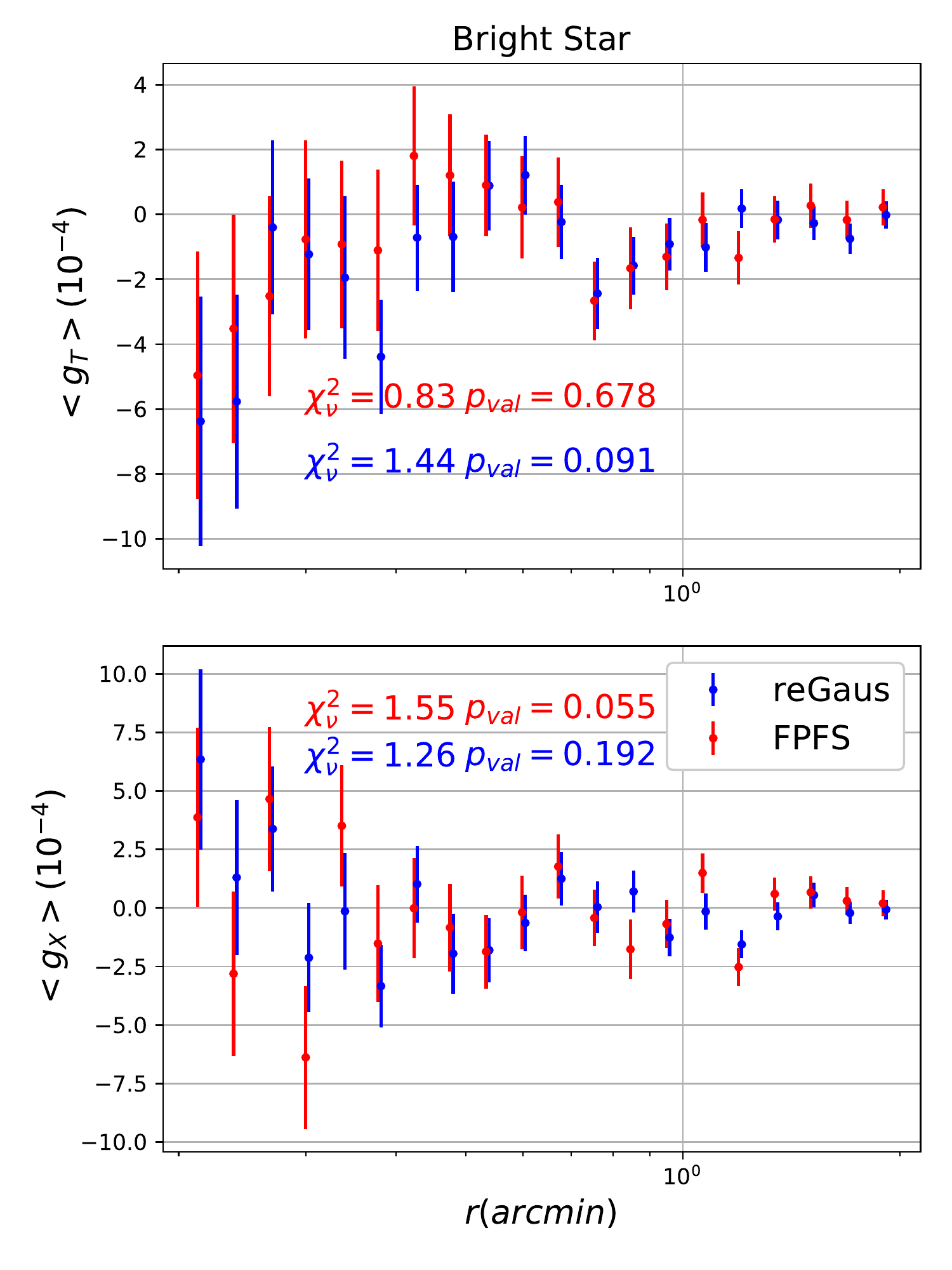}
\end{tabular}
\caption{The tangential (upper panels) and cross (lower panels) shear components measured by the FPFS shape catalog and the re-Gaussianization shape catalog around random points and bright stars. The red points are results for FPFS whereas the blue points are results for the re-Gaussianization shape catalog. The left panels show the results on random points. The number density of random points is $50~ \text{deg}^{-2}$. The right panels show the resuts on bright stars with $i$-band magnitude $\leq 22.5$. The reduced chi squres ($\chi_{\nu}^2$) and p-values are shown in each panel. Note that the reduced chi squres ($\chi_{\nu}^2$) and p-values are calculated without accounting for the correlation between data point at different radius bins. } \label{fig_ranBrighTest}
\end{figure*}
As mentioned in Section \ref{subsec:calib}, the differences between the simulation and observed data can lead to systematic bias when applying the calibration derived from the simulated data to the observed data and we refer to such bias as calibration residual.
Figure \ref{fig_EightHis} compares the overall number distributions between the simulation and observation over eight observables, which demonstrates that the differences between the simulation and observation are very small.

Here we quantify the amplitude of potential calibration residual caused by the differences between the simulation and the observation.
The original galaxy simulation, which is used to derive the calibration factor, is divided into $800$ subsamples with different input PSFs, noise variances, and galaxy shapes and this galaxy sample has a magnitude limit of $25.2$ \citep{Li18FPFS}.
We bootstrap these $800$ subsamples to simulate galaxies with different observational conditions and apply the shear estimator calibrated with the original simulation to the bootstrapped simulations with the additional $i$-band selections summarized in Table \ref{tab:source_selection}.
The remaining biases, including both the residual of multiplicative bias and the residual of additive bias, for each bootstrap realization are estimated and the distributions of the remaining biases are plotted in Figure \ref{fig_calibResiduals}.
As demonstrated in Figure \ref{fig_calibResiduals}, the centers of the distributions of the remaining biases are slightly offset from zero. However, the offsets are below $0.5\%$ for multiplicative bias and below $1\times 10^{-4}$ for additive bias, which are much smaller than the HSC first year science requirement. Furthermore, the probability distributions of the remaining biases shown in Figure \ref{fig_calibResiduals} also demonstrate that the remaining biases are within the HSC first year science requirement.

We note that such remaining biases not only include the calibration residual caused by the differences between the data used for calibration and the observed data but also they include the contribution of measurement error caused by the photon noise on the galaxy images of each subsamples. They also include the selection bias due to the $i$-band selections. Since we have not found any remaining biases exceeding the HSC first year science requirement, we leave the separation of calibration residual from the contributions of photon noise and selection bias to our future work.

{\redColor{Another possible cause of systematic bias is that additional cuts are used to select galaxy samples in the real observation. As shown in Table \ref{tab:source_selection} and Table \ref{tab:source_selection_photoz}, several selections based on multi-bands magnitudes and multi-bands SNRs are applied to the shape catalog to ensure the accuracy of photo-$z$ estimation.
Moreover, in the galaxy-galaxy lensing measurement performed in Section \ref{subsec:ESD}, an additional weight as function of photo-$z$ is added to optimize the excess surface density measurement.

To estimate the potential bias caused by such weighting and selection on photo-$z$, we further test the correlation between the calibration residual and the photo-$z$ by dividing our simulation into several COSMOS photo-$z$ \citep{COSMOS-photoz30Bands-Ilbert2009} bins and estimating the calibration residual in each bin.
As shown in the middle panel of  Figure \ref{fig_calibResiduals_mag_z_fwhm}, the calibration residuals weakly correlate with the COSMOS photo-$z$ but they are below the first year HSC requirement. Therefore, we conclude that the bias caused by the additional cuts and selection on photo-$z$ in the galaxy-galaxy lensing measurement should be within the first year HSC science requirement.

We also check the correlations between the calibration residual and other marginalized observables such as magnitude and seeing.
The left panel in Figure \ref{fig_calibResiduals_mag_z_fwhm} shows the calibration residuals for galaxies in different seeing bins and the right panel in Figure \ref{fig_calibResiduals_mag_z_fwhm} shows the calibration residuals for galaxies in different COSMOS F814W magnitude bins. We report that no potential bias beyond the first year HSC science requirement is discovered in these sanity tests.
}}

\section{Internal Null Tests}
\subsection{Mock Catalog}
\label{subsec:mockCatalog}
\begin{figure*}
\centering
\includegraphics[width=.9\textwidth]{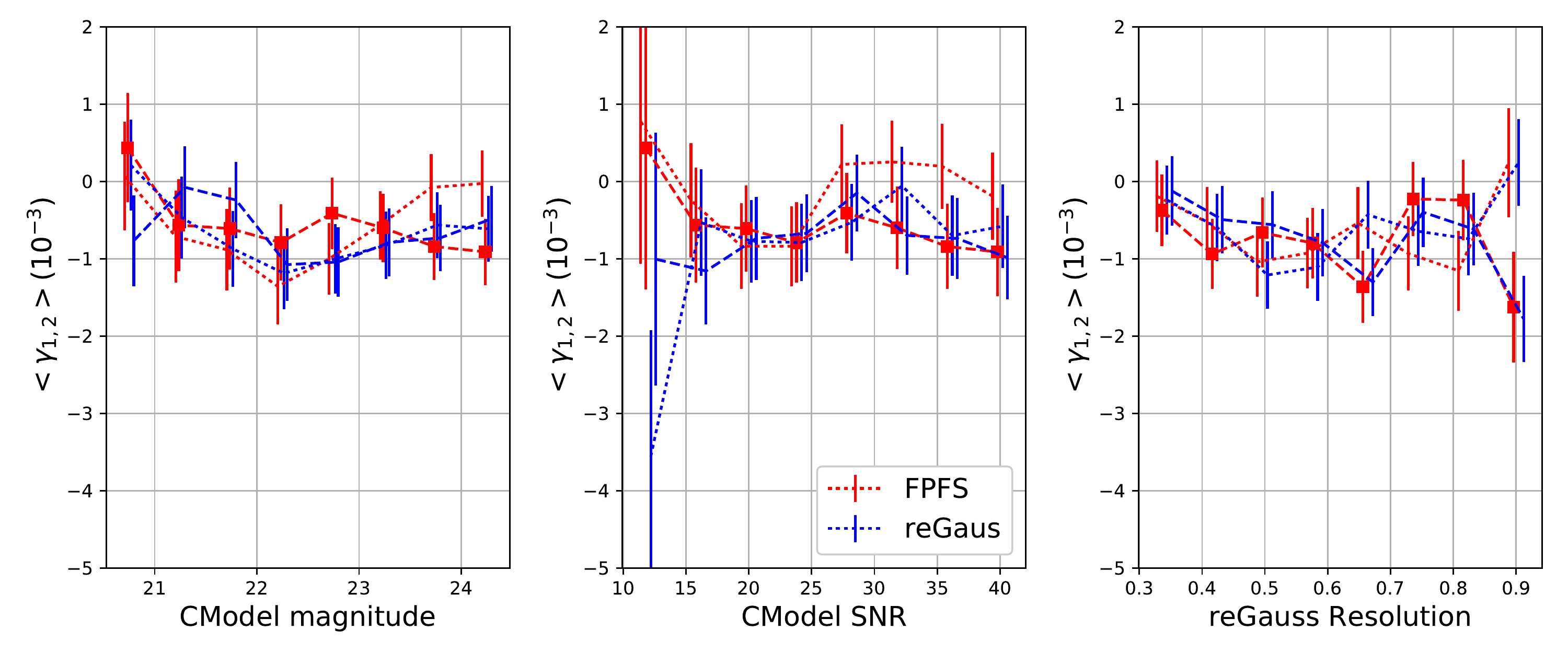}
\caption{The average $\gamma_1$ (dotted line) and $\gamma_2$ (dashed line) as functions of CModel magnitude (left panel), CModel S/N (middle panel), and CModel Resolution (right panel). The red lines are for the FPFS shape catalog and blue lines are for the re-Gaussianization shape catalog.} \label{fig_8BinsTest}
\end{figure*}
In order to check the significance of any non-zero values, it is necessary to accurately estimate errorbar of shear estimation which include both shape noise and cosmic variance. Error is dominated by shape noise on small scales due to the limited galaxy number whereas it is dominated by cosmic variance on large scales. Therefore, we construct mock catalogs to estimate error by adopting the real FPFS shape catalog and replacing the ellipticity of every individual galaxy with mock ellipticity \citep{HSC1-massMaps} which include cosmic shear signal from the ray-tracing simulation \citep{raytracingTakahashi2017}.

We keep the sky coordinates of all galaxies in mock catalogs the same as the real FPFS shape catalog described in Section \ref{subsec:FPFS_catalog}. The ellipticity of each galaxy is replaced with the mock values to simulate mock catalogs. The procedure of deriving the mock ellpticity is described as follows.

Firstly, every observed ellipticity is rotated with a random angle to eliminate the true shear signal and randomize the shape noise and photon noise. The rotated ellipticity is denoted as $e^{R}_{1,2}$.  Subsequently, we distort the rotated ellipticity to generate the mock ellipticity ($e^{M}_{1,2}$) by
\begin{equation}
e^{M}_{1,2}=e^{R}_{1,2}+\gamma_{1,2} R,
\end{equation}
where $\gamma_{1,2}$ is the cosmic shear obtained from all-sky weak lensing maps presented in \citet{raytracingTakahashi2017}. The cosmological model used for the all-sky simulation is from the best fitting result of the Wilkinson Microwave Anisotropy Probe (WMAP) nine-year data with $\Omega_M=0.279$, $\Omega_b=0.046$, $\Omega_\Lambda=0.721$, $b=0.7$, $n_s=0.97$, and $\sigma_8=0.82$ \citep{WMAP9th}. For each galaxy, we randomly assign its redshift following the MLZ photo-$z$ probability distribution function \citep{HSC1-photoz} of the galaxy. The shear values are obtained from the all-sky weak lensing maps at two adjacent redshift slices with linearly interpolation.

From these procedures, we create mock catalogs with different realizations of shape noise and cosmic shear. The distribution of the mock FPFS ellipticities are very similar to those of the real FPFS ellipticities.

\subsection{Null Tests}
\label{subsec:NullTest}
\begin{figure}
\centering
\includegraphics[width=.45\textwidth]{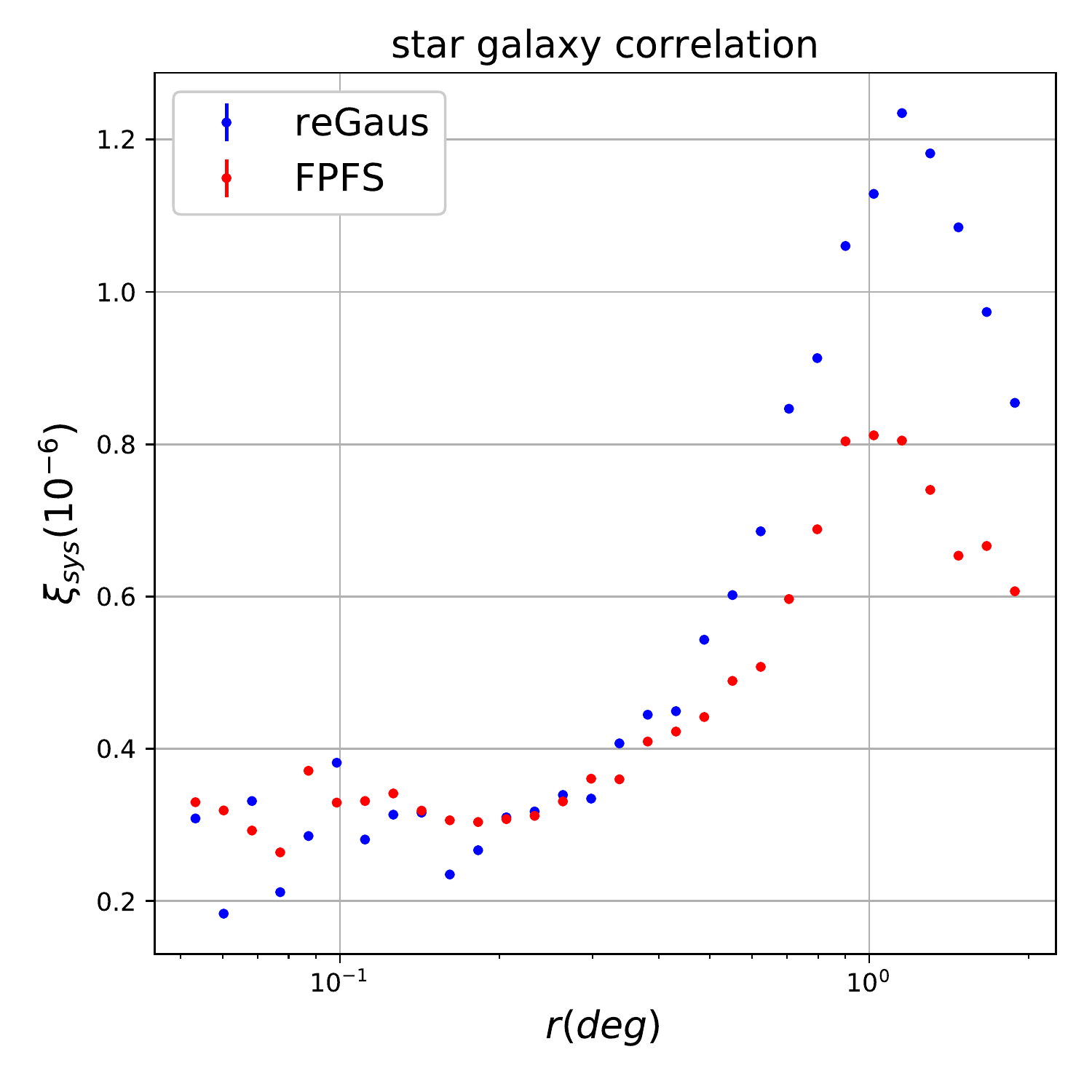}
\caption{Star galaxy correlation. The $x$-axis is the separtion angle between galaxies and stars. The $y$-axis is the PSF systematic parameter defined in equation (\ref{eq:PSF_systematics_define}).} \label{fig_starGalCor}
\end{figure}
We present a set of null tests \citep{HSC1-catalog} for the FPFS shape catalog and compare the results with the re-Gaussianization shape catalog. Note that all of the null tests adopt calibrations for multiplicative and additive biases. The results for the re-Gaussianization catalog can be found in \citet{HSC1-catalog}.

Figure \ref{fig_ranBrighTest} shows the tangential and cross shear components as functions of angular separations measured with the FPFS shape catalog and the re-Gaussianization shape catalog around random points and bright stars. The tests around random points extend to scales of $100$ acrmin with the intent to investigate systematic bias caused by the incomplete annuli on large-scales. The errorbars for the random point tests are estimated with the mock catalogs introduced in Section \ref{subsec:mockCatalog} since the errorbars are dominated by cosmic variance on large scales. The tests around bright stars focus on small scales within $2$ arcmin to investigate the possible systematic bias due to sky background mis-estimation near bright objects. The errorbars for the bright star tests are estimated using the mock catalogs with different realizations. The corresponding reduced $\chi^2$ and p-values for the fitting to zero signal are also shown in Figure \ref{fig_ranBrighTest}. It is necessary to note that when calculating the reduced $\chi^2$ and p-values, the correlations of error between different radius bins are not taken into account.  We do not find any significant detection of non-zero shear signal from these null tests. Moreover, we do not find significant difference between the results from two shape catalogs.

Figure \ref{fig_8BinsTest} shows the average of two shear components (namely $\left\langle g_{1,2} \right\rangle$) as functions of three $i$-band properties which are CModel $S/N$, CModel magnitude, and the re-Gaussianization resolution.
The errorbars in Figure \ref{fig_8BinsTest} are calculated with the mock catalogs introduced in Section \ref{subsec:mockCatalog}.
It is important to note that the errors in different property bins are correlated due to the cosmic variance such that the bin-to-bin correlation coefficients range from $0.3$ to $0.6$.
{\bColor In conclusion, we do not find strong dependence of the average shear values on these galaxy properties.
The averaged singal is persistently blow zero in most of the bins. Such negative signal is likely to be caused by cosmic variance.}

Figure \ref{fig_starGalCor} shows the results of the systematic test on star-galaxy shape correlations. As suggested by \citet{HSC1-catalog}, we define a parameter to quantify the performance of the PSF correction as
\begin{equation}\label{eq:PSF_systematics_define}
\xi_{\rm sys}=\frac{\left \langle \gamma_{*} \gamma_{\rm gal} \right \rangle}{\left \langle\gamma_{*} \gamma_{*} \right \rangle},
\end{equation}
where  $\gamma_{*}$ refers to the distortion measured from star image using the re-Gaussianization second order moments, and $\gamma_{\rm gal}$ is the distortion measured from galaxy images. We use the FPFS shape catalog and the re-Gaussianization shape catalog to measure $\gamma_{\rm gal}$ and determine $\xi_{\rm sys}$ as functions of the separation distance for two shape catalogs, respectively. As demonstrated in Figure \ref{fig_starGalCor}, while the overall amplitudes of $\xi_{\rm sys}$ measured from two shape catalogs are comparable, the systematic error for the FPFS shape catalog is slightly smaller than that of the re-Gaussianization shape catalog on large scales ($>0.5 \deg$).

\section{Galaxy-galaxy Lensing}
\subsection{Lens Catalog}
\label{subsec:lensCat}
\begin{figure}
\centering
\includegraphics[width=.45\textwidth]{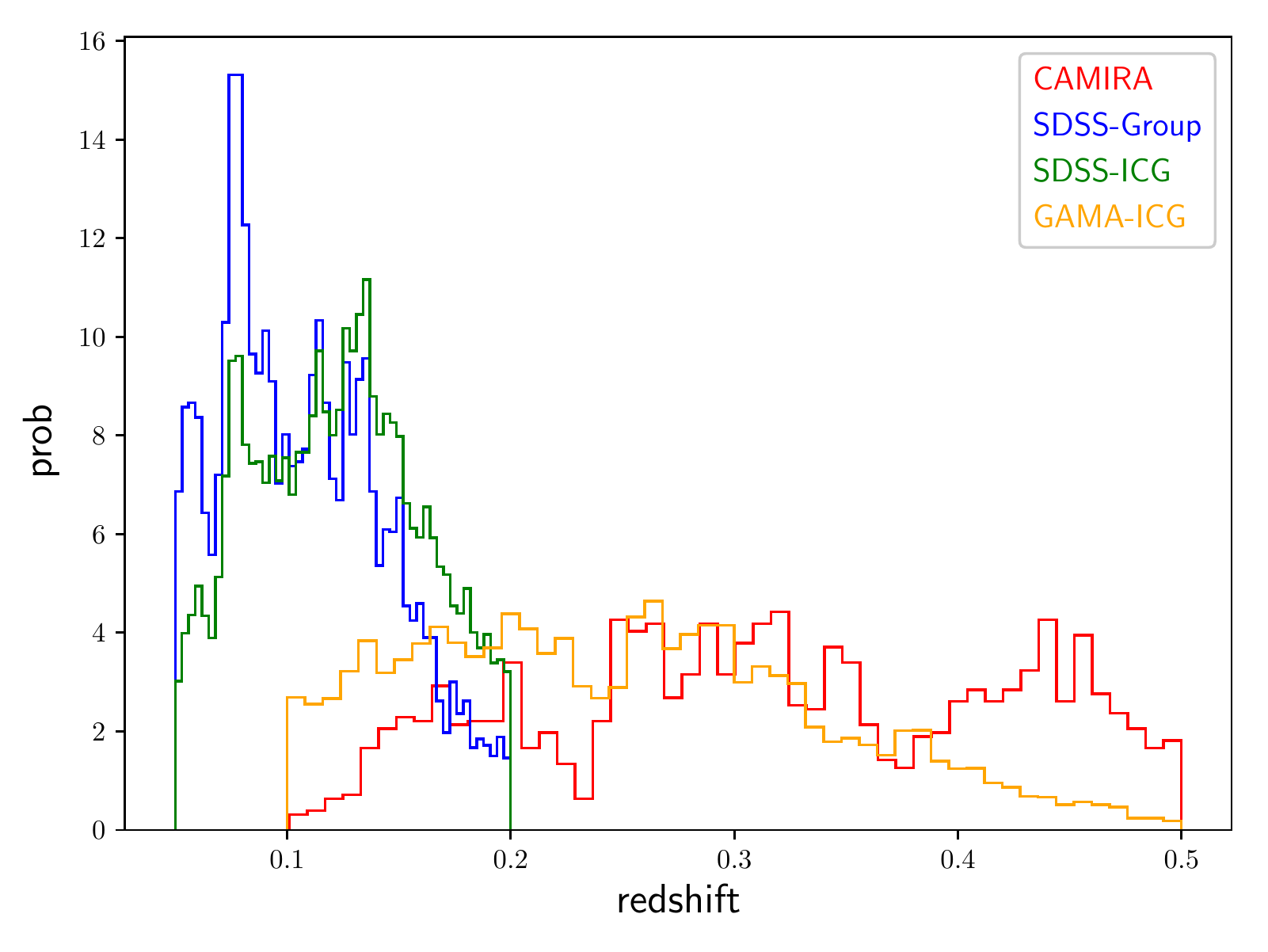}
\caption{The redshift distribution of lens catalogs used for galaxy-galaxy lensing. For SDSS group catalog and SDSS ICG catalog, lenses are selected with $0.05<z<0.2$. For GAMA ICG catalog and CAMIRA cluster catalog, lenses are selected with $0.1<z<0.5$.} \label{fig_lensZ}
\end{figure}
Observationally, a few different approaches have been adopted to identify a sample of galaxies with a high fraction of central galaxies in dark matter haloes, including red sequence cluster finder based on photometric surveys \citep[e.g.][]{CAMIRA,redMaPPer}, the construction of galaxy group catalogs based on spectroscopic data sets \citep[e.g.][]{SDSSgGYang}, and the selection of isolated galaxies which are isolated central galaxies \citep[ICG, e.g.][]{LBG-wang2012,LBG-wang2018}.
\begin{figure*}
	\centering
	\includegraphics[width=.9\textwidth]{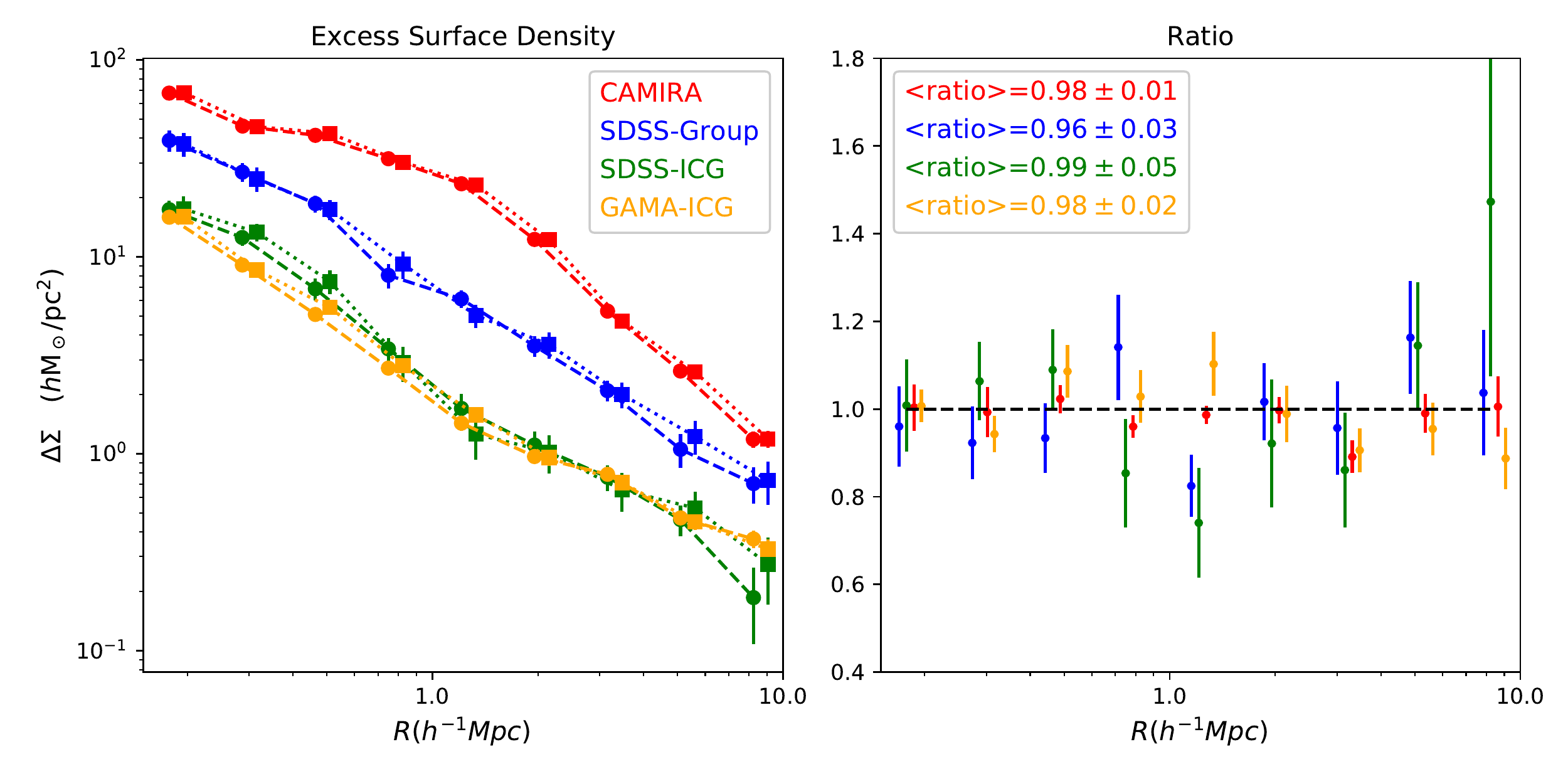}
	\caption{The left panel shows the ESDs measured by the re-Gaussianization shape catalog (dashed lines) and FPSF shape catalog (dotted lines) on different lens catalogs summarized in Section \ref{subsec:lensCat}. The right panel shows the ratio between the ESDs measrued by two shape catalogs. } \label{fig_gglens_internal}
\end{figure*}
We perform galaxy-galaxy lensing measurements on different lens catalogs, including HSC CAMIRA cluster catalog \citep{CAMIRA-HSC}, SDSS galaxy group catalog \citep{SDSSgGYang}, SDSS ICG catalog \citep{LBG-wang2012}, and GAMA \citep{GAMA-DR2011} ICG catalog. For each lens catalog, the galaxy-galaxy lensing analysis is conducted with two source galaxy catalogs, namely the FPFS shape catalog and the re-Gaussianization shape catalog to check the consistency between the results from two source catalogs. The selections of these four lens catalog are summarized below.

The HSC CAMIRA clusters \citep{CAMIRA-HSC} are constructed with the CAMIRA algorithm \citep{CAMIRA} from the Wide layer of the HSC first year data \citep{HSC1-data}. The CAMIRA algorithm is based on the Stellar Population Synthesis (SPS) model of \citet{SPT2003} to predict colors of red-sequence galaxies at a given redshift for an arbitrary set of bandpass filters. The algorithm applies additional calibration using spectroscopic galaxies to improve the accuracy of the SPS model prediction. Using the calibrated SPS model, CAMIRA algorithm computes the likelihood of each galaxy being in the red sequence as a function of redshift and create a richness map based on the likelihood. The CAMIRA clusters are detected as the peaks of the richness map, where the center of the cluster is defined as the center of the Brightest Cluster Galaxy (BCG).

The SDSS group catalog \citep{SDSSgGYang} is constructed with a modified version of the halo-based group finder developed by \citet{Yang-haloBasedgGFinder} from the NYU Value Added Galaxy Catalog \citep[NYU-VAGC;][]{NYU-VAGC}, which is based on the spectroscopic main galaxy sample from the fourth data release of the Sloan Digital Sky Survey \citep[SDSS/DR7;][]{SDSS-DR7}. This method finds potential group centers and its corresponding group members using Friend-Of-Friend (FOF) algorithm \citep{CDMSim-Davis1985} and determines a characteristic luminosity for each tentative group. With an assumed mass-to-light ratio, a tentative mass is assigned to each group and used to estimate the size and velocity dispersion of the underlying halo that hosts the group, which in turn is used to determine group memberships. With the updated group memberships, the group centers are also updated. This procedure is repeated until no further changes occur in group memberships.

Isolated central galaxies (ICG) are the brightest galaxy within a radius of $1$~Mpc in projected separation and within $1000$~ km/s along the line-of-sight. The parent sample used to select SDSS ICG is the NYU Value Added Galaxy Catalog \citep[NYU-VAGC;][]{NYU-VAGC}, which is based on the spectroscopic Main galaxy sample from the seventh data release of the Sloan Digital Sky Survey \citep[SDSS/DR7;][]{SDSS-DR7}. To avoid mistakenly selecting galaxies which have brighter physical companions, but the companions are missing spectroscopic redshifts due to the fiber collision, photo-$z$ probability distribution catalog \citep{photozCunha} is used to compensate the selection. Candidates are further rejected if they have apparent companions projected within $1$ Mpc and the photo-$z$ probability distribution of the companion is compatible with the spectroscopic redshift of the candidate \citep[see][for details]{LBG-wang2012}.

Compared with the SDSS Main galaxy sample, the Galaxy And Mass Assembly (GAMA) Survey \citep{GAMA-DR2011} is about two magnitudes deeper. The same selection criteria is applied to identify isolated central galaxies (ICG) from the public GAMA DR3 spectroscopy catalog \citep{GAMA-DR2018}. The selection steps are the same as for the SDSS Main galaxy sample, except that we do not apply any further selection using photo-$z$ probability distribution catalogs, as the effect of fiber collision in GAMA is much less severe.

Additional redshift cuts are applied to these lens catalogs. The redshift distributions of these four lens catalog after redshift cuts are shown in Figure \ref{fig_lensZ}.

\subsection{Excess Surface Density}
\label{subsec:ESD}
From weak lensing measurements, the average excess surface density (ESDs) of the lens samples described in Section \ref{subsec:lensCat} can be estimated.

The ESD for an isotropic lens system at redshift $z_l$ is defined as
\begin{equation}\label{eq:ESD-define}
\Delta\Sigma(R)=\Sigma(\leq R)-\Sigma (R),
\end{equation}
where $R$ is the radius to the center of the lens system, $\Sigma(\leq R)$ is the mean surface mass density inside a certain radius $R$, and $\Sigma (R)$ is the surface mass density at the radius $R$. The surface mass density refers to the line-of-sight projection of the mass density.

The lens system causes tangential shear to the background galaxies. The tangential shear at radius $R$ and redshift $z_s$ is
\begin{equation}
\gamma_T(R,z_l,z_s)=\Delta\Sigma (R) \Sigma_{cr}^{-1}(z_l,z_s),
\end{equation}
where the critical surface density is defined for a lens-source system as
\begin{equation}
\Sigma_{cr}^{-1}(z_l,z_s)=
\begin{cases}
\frac{4\pi G}{c^2}(1+z_l) \frac{\chi_{ls} \chi_l}{\chi_s} & (z_l\leq z_s)\\
0& (z_l > z_s)
\end{cases}.
\end{equation}
$\chi$ denotes the comoving distance, $G$ is the gravitational constant and $c$ is the speed of light. For simplicity, we denote $\Sigma^{-1}_{cr}(z_l,z_s)$ as $\Sigma^{-1}_{cr,ls}$. We use the PDF of the photometric redshifts ($P(z_s)$), estimated with MLZ algorithm \citep{HSC1-photoz}, to calculate the expectation of $\Sigma^{-1}_{cr,ls}$,
\begin{equation}
 \left\langle \Sigma_{cr,ls}^{-1} \right\rangle =\int_{z_l}^{+\infty}dz_s\Sigma_{cr}^{-1}(z_l,z_s)P(z_s).
\end{equation}
Since the uncertainties on the photo-$z$ measurements of background galaxies do not correlate with the ESDs of lens systems, the relation between the expectation of tangential shear and the expectation of ESD is
\begin{equation}
\left\langle\gamma_T(R,z_l,z_s)\right\rangle= \left\langle\Delta\Sigma (R)\right\rangle  \left\langle\Sigma_{cr}^{-1}(z_l,z_s)\right\rangle.
\end{equation}
In the galaxy-galaxy lensing analysis, source galaxies are further selected according to their redshift \citep{clusterLens-HSCSourceSelect2018}, which we summarized in Table \ref{tab:source_selection_photoz}.

We do not correct for the effect of photometric redshift bias on the lensing measurements (More et al. in prep) using COSMOS $30$-band photo-$z$ data as done in \citet{Mass-Calibration_actpol-HSC2019,CAMIRA-HSCY1-MassRichness-Murata2019}. Since the photometric bias is well below one percent for our selection criteria for lens and source galaxies \citep{Mass-Calibration_actpol-HSC2019}.

\begin{table*}[!]
\centering
\begin{tabular}{cl}
\hline\hline
photo-$z$ cut & descriptions\\
\hline
mlz\_std\_best$<$3      & The uncertainty of the photo-$z$ estimation\\
mlz\_conf\_best$>$0.13  & The confidence of the photo-$z$ estimation\\
mlz\_best$<$2.5         & The best estimation of photoz\\
$\int_{z_l+0.2}^{+\infty} P(z)>0.95$& P(z) cut defined in \citet{clusterLens-HSCSourceSelect2018}\\
\hline
\end{tabular}
\caption{Selection of source galaxies according to photo-$z$.}\label{tab:source_selection_photoz}
\end{table*}

The FPFS shear estimator divides the expectation of tangential ellipticities by the expectation of responses to estimate the tangential shear. The tangential ellipticity is defined as
\begin{equation}
e_{T} =-e_1 \cos(2\phi) -e_2 \sin(2\phi),
\end{equation}
where $\phi$ is the angular position of the source galaxy with respect to the centroid of lens system in polar coordinates. By stacking a large ensemble of lenses in the lens catalog, the stacked lens system is approximately isotropic and the expectation of stacked ESDs as function of radius is
\begin{equation}\label{eq:ESD_estimate_1}
\left\langle\Delta \Sigma (R) \right\rangle = \frac{ \sum_l \sum_s w_{ls} \left\langle\Sigma_{cr,ls}^{-1}\right\rangle^{-1} e_{T,s}}{\sum_l\sum_s (1+m_s)w_{ls} R_s },
\end{equation}
where $w_{ls}$ is the weight for each lens-source pair. Conventionally, the weight is set to
\begin{equation}\label{eq:lens-source_weight}
w_{ls}=\left\langle\Sigma_{cr}^{-1}(z_l,z_s)\right\rangle^2
\end{equation}
to optimize the estimation of excess surface density \citep{ESD18}. Substituting equation (\ref{eq:lens-source_weight}) into equation (\ref{eq:ESD_estimate_1}), the estimator of ESDs changes to
\begin{equation}\label{eq:ESD_estimate_final}
\left\langle\Delta \Sigma (R) \right\rangle = \frac{ \sum_l \sum_s \left\langle\Sigma_{cr,ls}^{-1}\right\rangle e_{T,s}}{\sum_l\sum_s (1+m_s)\left\langle\Sigma_{cr,ls}^{-1}\right\rangle^2 R_s }.
\end{equation}

Note that equation (\ref{eq:ESD_estimate_final}) is unbiased only if $\left\langle\Sigma_{cr}^{-1}(z_l,z_s)\right\rangle$ is not correlated with the ellipticities of galaxies. Here we assume that $\left\langle\Sigma_{cr}^{-1}(z_l,z_s)\right\rangle$ does not correlate with galaxy shapes without further validation since $\left\langle\Sigma_{cr}^{-1}(z_l,z_s)\right\rangle$ is estimated from the photo-$z$ PDF using multi-bands images and we do not have multi-bands image simulations to validate the assumption.
\break

\subsection{Results}
\label{subsec:gglens-results}
We compare the results of ESDs estimations between the re-Gaussianization shape catalog and the FPFS shape catalog on the lens catalogs summarized in Section \ref{subsec:lensCat}. The ESDs as functions of the distance to the center of the lenses for two shape catalogs are shown in the left panel of Figure \ref{fig_gglens_internal}. We focus on the radius larger than $0.1~\rm{Mpc}/h$, since the measurement is noisy due to the limited number of background source galaxies on the smaller scale. The error bars for the ESDs are estimated with $100$ realizations of galaxy shape catalogs with randomly rotated shapes.

The right panel of Figure \ref{fig_gglens_internal} shows the ratio between the measurements from two shape catalogs, where the ratio is defined as
\begin{equation}\label{eq:ESD-ratio}
ratio=\frac{\left\langle\Delta\Sigma_{FP}\right \rangle}{\left\langle\Delta\Sigma_{RG}\right\rangle},
\end{equation}
where $\left\langle\Delta\Sigma_{FP}\right\rangle$ and $\left\langle\Delta\Sigma_{RG}\right\rangle$ are the ESDs measured with the FPFS shape catalog and the re-Gaussianization shape catalog, respectively.
When calculating the error bars of the ratio, it is necessary to account for the correlation between the ESDs measured with two shape catalogs \citep{Mass-Calibration_actpol-HSC2019}.
The correlation is also estimated from the $100$ realizations of galaxy shape catalogs with randomly rotated shapes for each lens catalog.
{\oColor{The ratio of the last radius bin for SDSS-ICG lens sample is biased high since the error is comparable to the denominator in equation (\ref{eq:ESD-ratio}).}}
We report that no difference beyond the statistical uncertainty has been found between these two shape catalogs.

\section{Mass Map}
\label{sec:kappaMap}

\subsection{Kaiser-Squires reconstruction}
\label{subsec:kappaMap-KS}
In addition to measuring the ESDs around lens catalogs, we also construct the projected mass distribution from the observed shear map \citep{massMap-KS1993}.

We smooth the shear field with Gaussian filter as suggested by \citet{HSC1-massMaps}. The Gaussian smoothing filter is defined as
\begin{equation}
W(\vec{\theta}) = \frac{1}{\pi \theta_s^2}\exp\left(-\frac{|\vec{\theta}|^2}{\theta_s^2}\right    ),
\end{equation}
where the smoothing scale $\theta_s$ is set to $2$ arcmin in the following analysis. The smoothed two components of shear field are
\begin{equation}
\gamma_{1,2} (\vec{\theta})=\frac{\sum_i e_{1,2}(\vec{\theta_i}) W(|\vec{\theta}-\vec{\theta}_i|) }{\sum_i (1+\hat{m}(\vec{\theta_i}))R(\vec{\theta_i})  W(|\vec{\theta}-\vec{\theta}_i|)},
\end{equation}
where $e_{1,2}(\vec{\theta_i})$, $\hat{m}(\vec{\theta_i})$, and $R(\vec{\theta_i})$ are two components of ellipticity, multiplicative bias, and average response for the galaxy at position $\vec{\theta_i}$, respectively. The smoothed shear fields is pixelized on a regular grid adopting a flat-sky approximation. The pixel size is $0.5$ arcmin.

Subsequently, the shear field is converted to the convergence field via \citep{massMap-KS1993}
\begin{equation}\label{eq:KS_real}
\kappa(\vec{\theta})=\frac{1}{\pi} \int d^2\theta' \frac{\gamma_t(\vec{\theta'}|\vec{\theta})}{|\vec{\theta}-\vec{\theta'}|^2},
\end{equation}
where $\hat{\gamma}_t(\vec{\theta'}|\vec{\theta})$ is a tangential shear at position $\vec{\theta'}$ computed with respect to the reference position $\vec{\theta}$. The shear-to-convergence relationship is a convolution in two dimensional angular plane. Such convolution is computed in Fourier space with the Fast Fourier Transform (FFT) to reduce the computational time. Shear fields are padded with zero beyond their boundary to avoid the periodic boundary condition assumed in FFT. The complex shear field is denoted as $\gamma(\vec{\theta})=\gamma_1(\vec{\theta})+i\gamma_2(\vec{\theta})$. The Fourier transform of complex shear field and complex kappa field is denoted as $\tilde{\gamma}(\vec{l})$ and $\tilde{\kappa}(\vec{l})$, respectively. Equation (\ref{eq:KS_real}) can be expressed in Fourier space as
\begin{equation}
\tilde{\kappa}(\vec{l})=\pi^{-1}\tilde{\gamma}(\vec{l})\tilde{D}^{*}(\vec{l}) ~~ for ~~\vec{l}\neq\vec{0},
\end{equation}
where $\tilde{D}(\vec{l})$ is the Fourier transform of the convolution kernel in equation (\ref{eq:KS_real})
\begin{equation}
\tilde{D}(\vec{l})=\pi \frac{l_1^2-l_2^2+2il_1l_2}{|\vec{l}|^2}.
\end{equation}
\break
The mass map in configuration space $(\kappa(\vec{\theta}))$ is then reconstructed by inverse Fourier transforming $\tilde{\kappa} (\vec{l})$.
Note that the real part of the reconstructed mass map is referred to as an E-mode mass map, whereas the imaginary part of the reconstructed mass map is referred to as a B-mode mass map which is used to check for certain types of residual systematics in weak lensing measurements.

A `sigma map' of the convergence field is constructed as follows. We randomly rotate the ellipticity of every individual galaxy to randomize the shape catalog and construct the mass map with the randomized shape catalog. Such procedure is repeated to create $100$ random mass maps with different realizations of randomized shape catalogs. Then a standard deviation is calculated for each pixel from the 100 random mass maps to construct a `sigma map' which shows the spatial variation for the statistical noise of the reconstructed mass map.
\subsection{Results}
\label{subsec:kappaMap-res}
\begin{figure*}
	\centering
	\includegraphics[width=.95\textwidth]{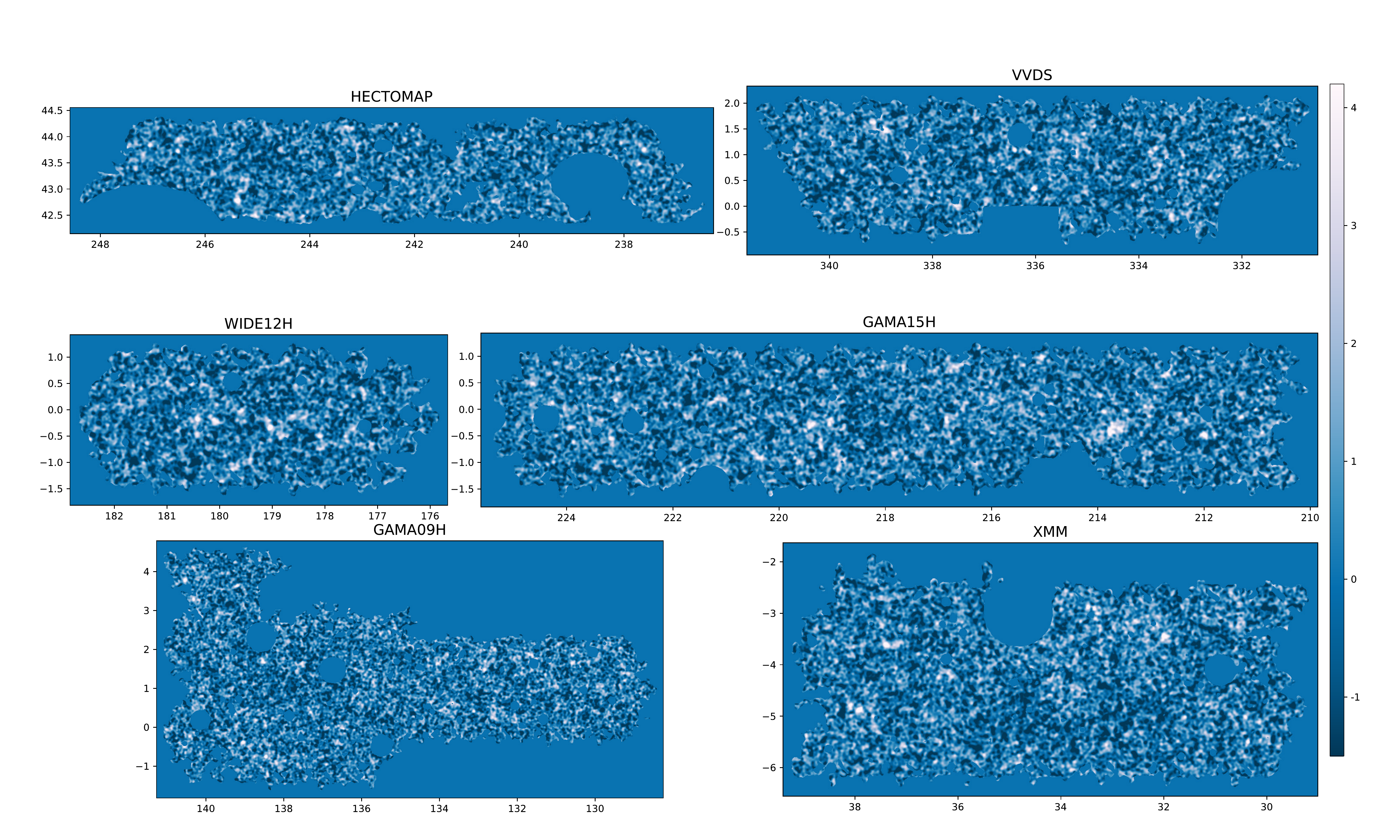}
	\caption{The convergence S/N maps for six fileds of the HSC first year data.} \label{fig_kSnrMap}
\end{figure*}
Figure \ref{fig_kSnrMap} shows the S/N maps for each fields of the HSC first year data. S/N of convergence is defined as a convergence divided by a `sigma' of statistical noise. Therefore, the S/N maps are calculated by dividing the mass maps with the `sigma maps'.

\begin{figure}
	\centering
	\includegraphics[width=.45\textwidth]{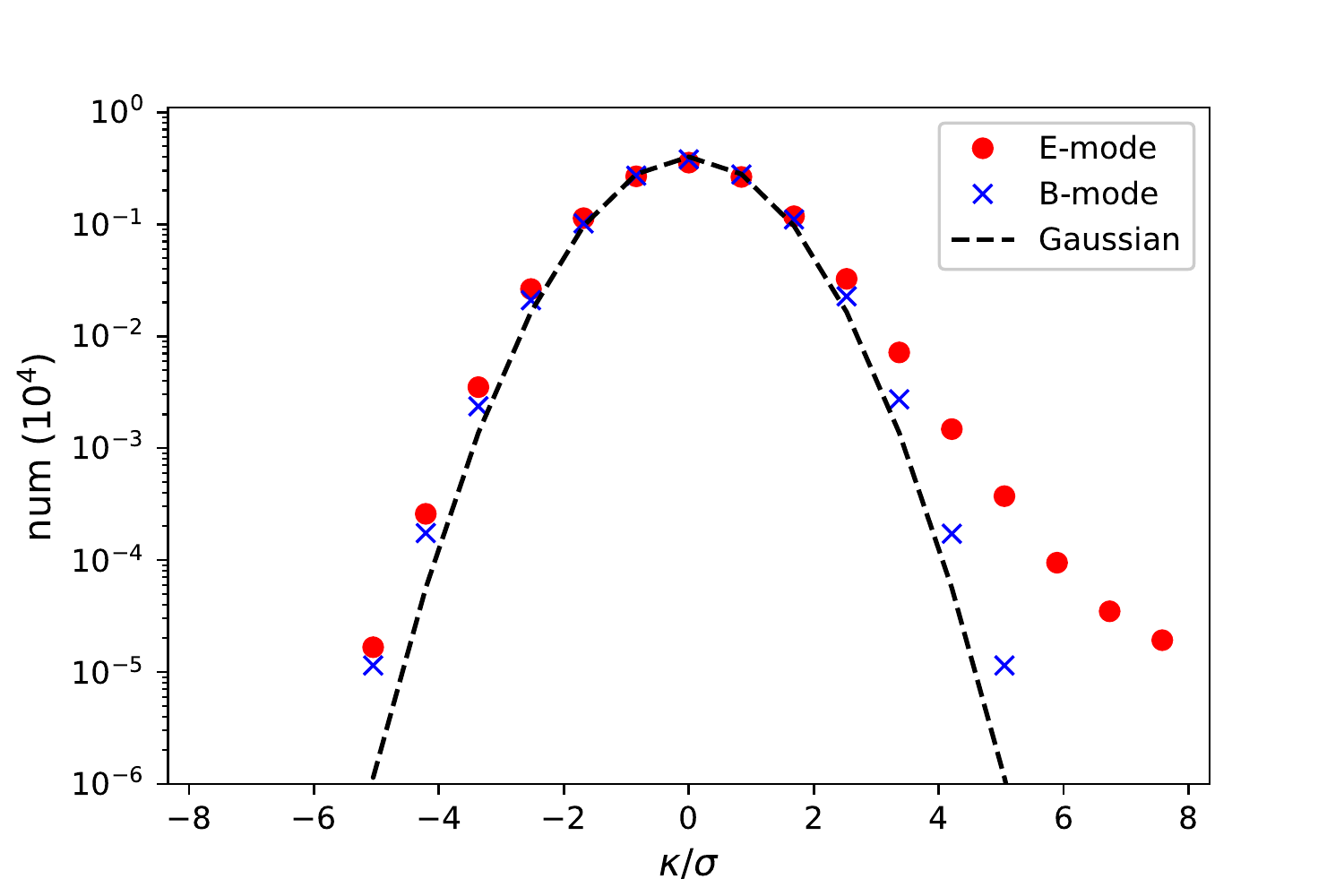}
	\caption{The probability density function (PDF) for the pixel values of E(B)-mode convergence S/N maps. The red points are for E-mode and the blue points are for B-mode. The dashed line is the PDF of a Gaussian distribution. } \label{fig_kSnrPDF}
\end{figure}
Figure \ref{fig_kSnrPDF} shows the probability density function (PDF) for the pixel values of E-mode and B-mode convergence S/N maps. The E-mode PDF significantly deviates from a Gaussian distribution with a high positive S/N tail. On the other hand, for the B-mode PDF, we do not find any tail from both the negative and positive ends. We note that the B-mode PDF is slightly deviated from a Gaussian distribution, which could be caused by the finite survey area and the irregular masks \citep[see][]{HSC1-massMaps}.

\section{Conclusion}
In this paper we apply the FPFS shear estimator to the first year data of the HSC survey. The FPFS shear estimator measures shapelet modes from Fourier power function of galaxy images after deconvolving PSF in Fourier space and uses four shapelet modes to construct ellipticity and shear response.

We perform both tests with external image simulation and null tests to validate the accuracy of the FPFS shape catalog. The external galaxy image simulation combined with the star/PSF images of the HSC first year data is used to demonstrate that the biases caused by the PSF model residual are within the HSC first year science requirement. Moreover, we use the HSC-like image simulation to demonstrate that the biases, which are caused by the difference between the simulated data used for calibration and the observed data to which the calibration is applied, are also much smaller than the HSC first year science requirement.
The internal tests are conducted within the HSC data set. The internal test results of the FPFS shape catalog is compared with those of the re-Gaussianization shape catalog. We find that the results of two catalogs are consistent with each other and there are no remaining systematic biases beyond the HSC first year science requirement.

The FPFS shape catalog is used to measure the excess surface density of several lens catalogs and the results are compared with those of the re-Gaussianization shape catalog. We find that the results from the two catalogs are consistent with each other within the statistical errors. Also, we apply the FPFS shape catalog to mass map reconstructions and we find no $B$-mode signals from the reconstructed mass maps.

Even though the assumptions behind these two shear estimators are different, both of these methods use the same image simulation to calibrate shear bias so the systematic uncertainties of these two shape catalogs are not strictly independent. Nevertheless, we note that the procedures of modeling and calibration of systematic biases are also different among these two methods. The re-Gaussianization shape catalog uses the image simulation to determine optimal weight and response as function of $S/N$ and resolution and calibrates noise bias, model bias, weight bias, selection bias and blending bias using the simulation. On the other hand, FPFS only uses the image simulation to calibrate blending bias. Therefore, we expect that cross comparisons of any scientific
results between these two catalogs on the observed data are still valuable.

\section*{Acknowledgements}
We thank Jun Zhang and Rachel Mandelbaum for useful comments.

This work was supported by Global Science Graduate Course (GSGC) program of University of Tokyo and JSPS KAKENHI (JP19J22222, JP15H05892, JP18K03693).

The Hyper Suprime-Cam Subaru Strategic Program (HSC-SSP) is led by the astronomical communities of Japan and Taiwan, and Princeton University.  The instrumentation and software were developed by the National Astronomical Observatory of Japan (NAOJ), the Kavli Institute for the Physics and Mathematics of the Universe (Kavli IPMU), the University of Tokyo, the High Energy Accelerator Research Organization (KEK), the Academia Sinica Institute for Astronomy and Astrophysics in Taiwan (ASIAA), and Princeton University.  The survey was made possible by funding contributed by the Ministry of Education, Culture, Sports, Science and Technology (MEXT), the Japan Society for the Promotion of Science (JSPS),  (Japan Science and Technology Agency (JST),  the Toray Science Foundation, NAOJ, Kavli IPMU, KEK, ASIAA,  and Princeton University.

Funding for the Sloan Digital Sky Survey IV has been provided by the Alfred P. Sloan Foundation, the U.S. Department of Energy Office of Science, and the Participating Institutions. SDSS-IV acknowledges support and resources from the Center for High-Performance Computing at the University of Utah. The SDSS web site is www.sdss.org.

SDSS-IV is managed by the Astrophysical Research Consortium for the Participating Institutions of the SDSS Collaboration including the Brazilian Participation Group, the Carnegie Institution for Science, Carnegie Mellon University, the Chilean Participation Group, the French Participation Group, Harvard-Smithsonian Center for Astrophysics, Instituto de Astrof\'isica de Canarias, The Johns Hopkins University, Kavli Institute for the Physics and Mathematics of the Universe (IPMU), University of Tokyo, the Korean Participation Group, Lawrence Berkeley National Laboratory,
Leibniz Institut f\"ur Astrophysik Potsdam (AIP),
Max-Planck-Institut f\"ur Astronomie (MPIA Heidelberg),
Max-Planck-Institut f\"ur Astrophysik (MPA Garching),
Max-Planck-Institut f\"ur Extraterrestrische Physik (MPE),
National Astronomical Observatories of China, New Mexico State University,
New York University, University of Notre Dame,
Observat\'ario Nacional / MCTI, The Ohio State University,
Pennsylvania State University, Shanghai Astronomical Observatory,
United Kingdom Participation Group,
Universidad Nacional Aut\'onoma de M\'exico, University of Arizona,
University of Colorado Boulder, University of Oxford, University of Portsmouth,
University of Utah, University of Virginia, University of Washington, University of Wisconsin,
Vanderbilt University, and Yale University.

GAMA is a joint European-Australasian project based around a spectroscopic campaign using the Anglo-Australian Telescope. GAMA is funded by the STFC (UK), the ARC (Australia), the AAO, and the participating institutions. The GAMA website is http://www.gama-survey.org/.

\bibliographystyle{aasjournal}
\bibliography{weak_lensing,lensCat,cosmoSim,cmbObs,gglens,shearCor,massMap,other}

\appendix
\section{Correlation between the additive bias and PSF ellipticity}
\label{Appendix:corCeP}
In \citet[][Figure 10]{Li18FPFS}, we checked the correlation between the additive bias and the PSF ellipticity for isolated galaxies without considering the PSF model residual. We concluded that the amplitude of fractional additive bias is below $0.5\%$.
Here we further check the correlation between the additive bias and the ellipticity of PSF taking account of blending and PSF residual.
\begin{figure}
    \centering
    \includegraphics[width=.95\textwidth]{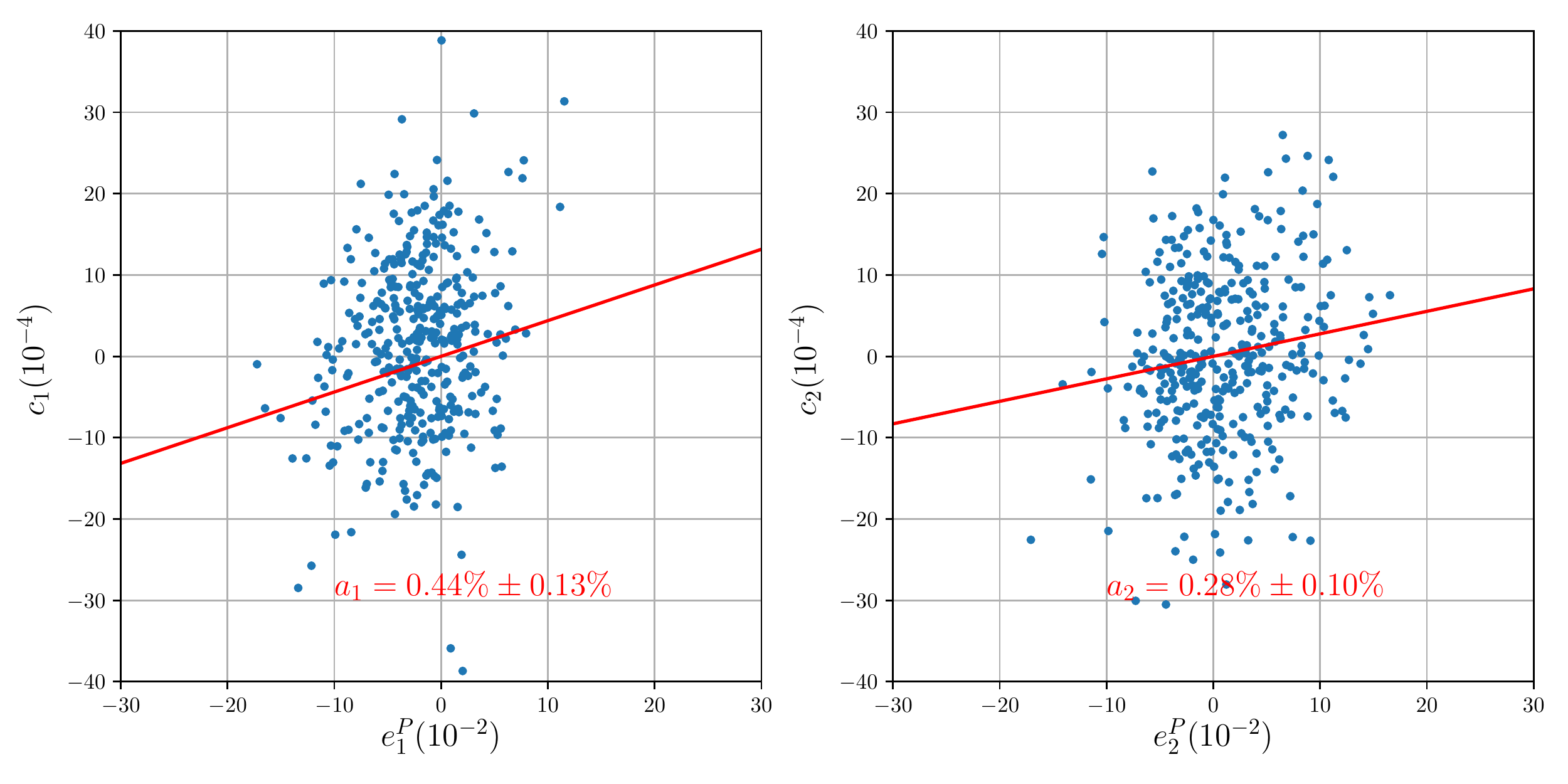}
    \caption{The correlation between the additive bias and the PSF ellipticity for sample 4 (with blending) of the HSC-like Great3 simulation \citep{HSC1-GREAT3Sim}. Each point is the result for one subfield in the simulation. The solid lines in the two panels show the fitting relations between the additive bias and PSF ellipticity. The $x$-axes are two components of the PSF ellipticity and the $y$-axes are two components of the additive bias.}
    \label{fig_psfAnBlend}
\end{figure}{}

Figure \ref{fig_psfAnBlend} shows the correlation between the additive bias and the PSF ellipticity for sample 4 of the HSC-like Great3 simulation \citep{HSC1-GREAT3Sim}. Sample 4 in \citet{HSC1-GREAT3Sim} is realistic galaxy sample containing both isolated galaxies and blended galaxies. Each scatter point is the result for one subfield in the simulation. The fractional additive bias is shown in Figure \ref{fig_psfAnBlend}, the amplitude of which is below $0.5\%$.

\begin{figure}
    \centering
    \includegraphics[width=.95\textwidth]{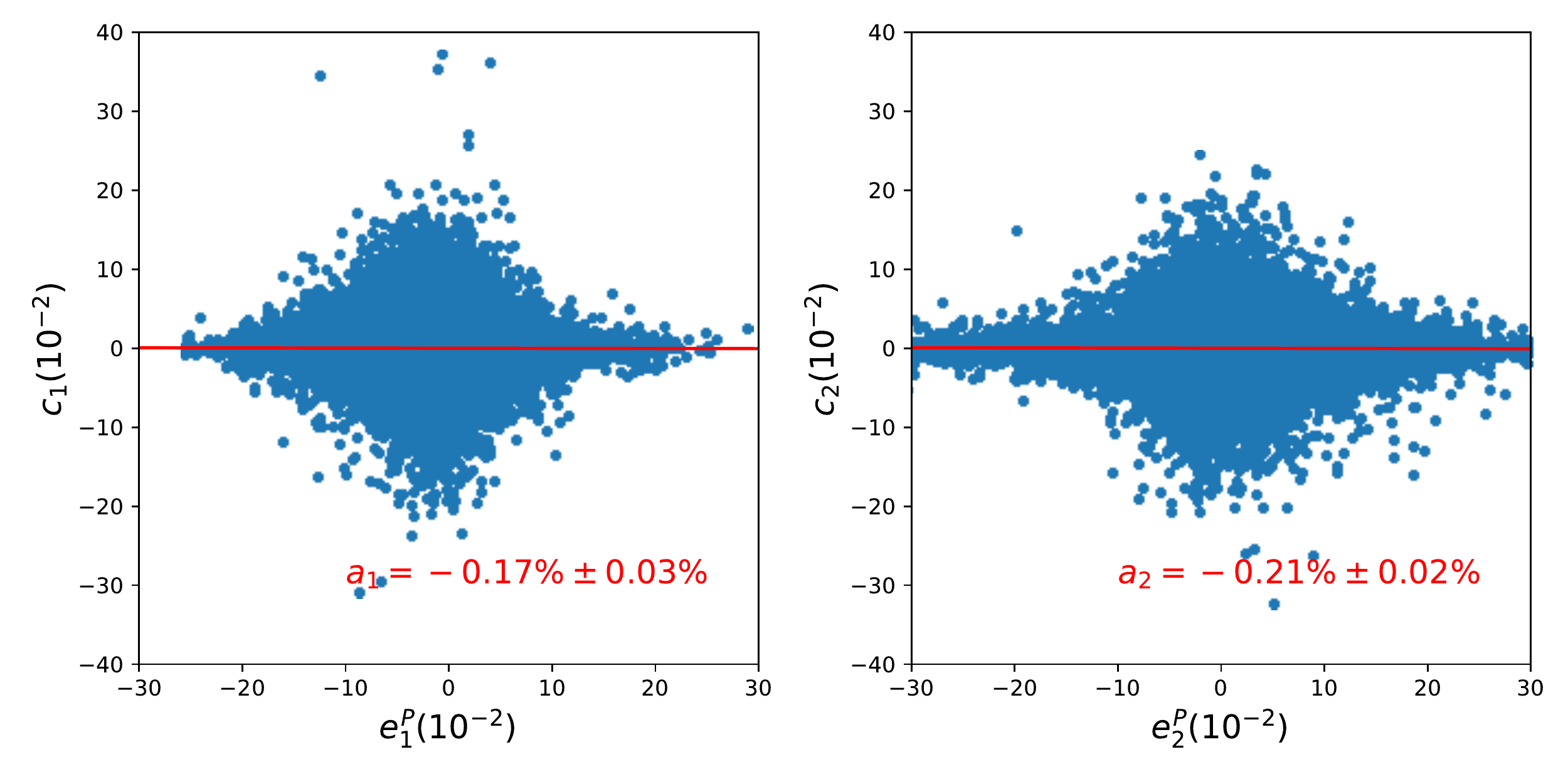}
    \caption{The correlation between the additive bias and the PSF ellipticity for simulation described in Section \ref{subsec:psfTest}. Each point is the result for one star in the star sample selected in Section \ref{subsec:psfTest}. The solid lines in the two panels show the fitting relations between the additive bias and PSF ellipticity. The $x$-axes are two components of the PSF ellipticity and the $y$-axes are two components of the additive bias.}
    \label{fig_psfAnPsfRes}
\end{figure}{}
However, sample 4 of the HSC-like Great3 simulation does not include the PSF model residual. Figure \ref{fig_psfAnPsfRes} shows the correlation between the additive bias and the PSF ellipticity for the simulation described in Section \ref{subsec:psfTest}. Each point is the result for one star in the star sample selected in Section \ref{subsec:psfTest}. The fractional additive bias is also shown in Figure \ref{fig_psfAnPsfRes}, the amplitude of which is below $0.5\%$.

\section{Cosmic shear measurement}
\label{Appendix:cosmic shear}
Here we describe the procedure of using the FPFS shape catalog to measure the shear-shear correlation functions. As shown in equation (\ref{eq:shearCor_psfres}), two components of the shear-shear correlation function are
\begin{equation}
\hat{\xi}^\gamma_{\pm} (|\vec{\theta}|) =\left\langle \hat{\gamma}_t (\vec{\theta}_i) \hat{\gamma}_t (\vec{\theta}_j)\right\rangle \pm \left\langle \hat{\gamma}_\times (\vec{\theta}_i) \hat{\gamma}_\times (\vec{\theta}_j)\right\rangle,
\end{equation}
where $\hat{g}_t$ and $\hat{g}_\times$ are the tangential component and cross component with the reference to the separation, and the expectations are taken over pairs of galaxies with angular separation $|\vec{\theta}| = |\vec{\theta}_i -\vec{\theta}_j |$ within an interval $\Delta |\vec{\theta}|$ around $|\vec{\theta}|$.

For each source pairs, two compnents of galaxy ellipticity are decomposed into tangential ($e_t$) and cross ($e_\times$) components. Subsequently, we have
\begin{equation}
\begin{split}
\left\langle \hat{\gamma}_t (\vec{\theta}_i) \hat{\gamma}_t (\vec{\theta}_j)\right\rangle&=\frac{\sum_{i,j} e_{t} (\vec{\theta}_i) e_{t} (\vec{\theta}_j)}{\sum_{i,j} \left[1+m (\vec{\theta}_i)\right] R(\vec{\theta}_i) \left[1+m (\vec{\theta}_j )\right] R(\vec{\theta}_j)},\\
\left\langle \hat{\gamma}_\times (\vec{\theta}_i) \hat{\gamma}_\times (\vec{\theta}_j)\right\rangle&=\frac{\sum_{i,j} e_{\times} (\vec{\theta}_i) e_{\times} (\vec{\theta}_j)}{\sum_{i,j} \left[1+m (\vec{\theta}_i)\right] R(\vec{\theta}_i) \left[1+m (\vec{\theta}_j )\right] R(\vec{\theta}_j)}.
\end{split}
\end{equation}

\end{document}